\documentclass[twocolumn,pra,aps,superscriptaddress,floatfix]{revtex4}
\usepackage{amssymb}
\usepackage{amsfonts}
\usepackage{amsmath}
\usepackage{graphicx}

\begin{document}

\title{A solvable class of non-Markovian quantum multipartite dynamics}
\author{Adri\'an A. Budini}
\affiliation{Consejo Nacional de Investigaciones Cient\'{\i}ficas y T\'{e}cnicas
(CONICET), Centro At\'{o}mico Bariloche, Avenida E. Bustillo Km 9.5, (8400)
Bariloche, Argentina, and Universidad Tecnol\'{o}gica Nacional (UTN-FRBA),
Fanny Newbery 111, (8400) Bariloche, Argentina}
\author{Juan P. Garrahan}
\affiliation{School of Physics and Astronomy, University of Nottingham, Nottingham, NG7
2RD, UK}
\affiliation{Centre for the Mathematics and Theoretical Physics of Quantum
Non-Equilibrium Systems, University of Nottingham, Nottingham, NG7 2RD, UK}
\date{\today }
\begin{abstract}
We study a class of multipartite open quantum dynamics for systems of
arbitrary number of qubits. The non-Markovian quantum master equation can
involve arbitrary single or multipartite and time-dependent dissipative
coupling mechanisms, expressed in terms of strings of Pauli operators. We
formulate the general constraints that guarantee the complete positivity of
this dynamics. We characterize in detail underlying mechanisms that lead to
memory effects, together with properties of the dynamics encoded in the
associated system rates. We specifically derive multipartite ``eternal''
non-Markovian master equations that we term hyperbolic and trigonometric due
to the time dependence of their rates. For these models we identify a
transition between positive and periodically divergent rates. We also study
non-Markovian effects through an operational (measurement-based) memory
witness approach.
\end{abstract}

\maketitle

\section{Introduction}

In the theory of open quantum systems, the formulation of quantum Markovian
master equations is completely determined by the theory of quantum
semigroups \cite{alicki}. In contrast, the study of non-Markovian memory
effects presents two problems. The first one is that the most general
structure of a quantum master equation that captures memory effects, and at
the same time is consistent with the completely positive (CP) condition of
the solution map \cite{breuerbook,vega,wiseman}, is not known. The second
one is that different inequivalent memory witnesses can be used to define
and measure non-Markovian effects \cite{BreuerReview,plenioReview}.

The first problem has been known for many years. In fact, arbitrary
non-Markovian quantum master equations may lead to unphysical solutions \cite%
{wilkie,barnett,budini,cresserJD} where the average state (the density
matrix) being not positive definite. For tackling this issue a broad class
of phenomenological and theoretical approaches has been formulated \cite%
{vega}, dealing with both time-convoluted and convolutionless master
equations \cite{LocalNonLocal}. Examples include the dynamics induced by
stochastic Hamiltonians defined by non-white noises \cite{GaussianNoise},
phenomenological single memory kernels \cite%
{shabani,petruccioneLidarEq,salo,kossaDariusz}, interaction with incoherent
degrees of freedom \cite%
{lindbladrate,PostMarkovian,boltzman,pekola,megier,maximal} and arbitrary
ancilla systems \cite{swf,hush}, related quantum collisional models \cite%
{embedding,collisionVacchini,ciccarello,palmaMultipartito,strunz,portugal,brasilCollisional,brasil}%
, quantum generalizations of semi-Markov processes \cite{Semi,andrez}, and
random unitary dynamics \cite{wudarski,Polonia}, together with some exact
derivations from underlying (microscopic or effective) unitary dynamics \cite%
{TwoQubits,exactDecayTLS,DivergingRatesJCModel,deltaCorrelated,ferialdi,exactChina,plenio,additivity,smirne}%
.

Despite these advances \cite%
{wilkie,barnett,budini,cresserJD,LocalNonLocal,GaussianNoise,shabani,petruccioneLidarEq,salo,kossaDariusz,lindbladrate,PostMarkovian,boltzman,pekola,megier,maximal,embedding,collisionVacchini,ciccarello,palmaMultipartito,strunz,portugal,brasilCollisional,brasil,Semi,andrez,wudarski,Polonia,TwoQubits,exactDecayTLS,DivergingRatesJCModel,deltaCorrelated,ferialdi,exactChina,plenio,additivity,smirne,swf,hush}
most studies of non-Markovian evolutions are restricted in general to single
or bipartite systems. In fact, in general checking the CP condition of the
dynamics is a non trivial task, whose difficulty in turn increases with the
system's Hilbert space dimension. However, quantum information intrinsically
requires multipartite processing, and as a consequence the formulation of
multipartite non-Markovian dynamics is of interest from both theoretical and
practical points of view.

Our main goal in this paper is to formulate and study a class of solvable
multipartite non-Markovian master equations. The class of systems we
consider are defined in terms an arbitrary number $N$ of qubits, whose
interaction with the environment can be taken into account through arbitrary
Pauli channels. The evolution of the system's density matrix $\rho _{t}$ is
given by the time-local master equation \ $(d/dt)\rho _{t}=\mathcal{L}[\rho
_{t}],$ where the generator of the evolution has the general structure%
\begin{eqnarray}
\ \mathcal{L}[\bullet ]\! &=&\!\!\!\sum_{\substack{ i=1,\cdots N \\ \alpha
=x,y,z}}\!\!\Gamma _{i}^{\alpha }(t)(\sigma _{i}^{\alpha }\bullet \sigma
_{i}^{\alpha }-\bullet )\!  \label{Motivation} \\
&&\!\!\!\!+\!\!\!\!\sum_{\substack{ i=1,\cdots N \\ \alpha ,\beta =x,y,z}}%
\!\!\Gamma _{i}^{\alpha \beta }(t)(\sigma _{i}^{\alpha }\sigma _{i+1}^{\beta
}\bullet \sigma _{i+1}^{\beta }\sigma _{i}^{\alpha }-\bullet )  \notag \\
&&\!\!\!\!+\!\!\!\!\sum_{\substack{ i=1,\cdots N \\ \alpha ,\beta ,\gamma
=x,y,z}}\!\!\!\!\Gamma _{i}^{\alpha \beta \gamma }(t)(\sigma _{i}^{\alpha
}\sigma _{i+1}^{\beta }\sigma _{i+2}^{\gamma }\bullet \sigma _{i+2}^{\gamma
}\sigma _{i+1}^{\beta }\sigma _{i}^{\alpha }-\bullet )  \notag \\
&&\!\!\!\!+\cdots .  \notag
\end{eqnarray}%
Here, $\sigma _{i}^{\alpha }$ is the $\alpha $-th Pauli operator $(\alpha
=x,y,z)$ acting on qubit $i$, while $\Gamma _{i}^{\alpha \cdots \beta }(t)$
define local and bipartite time-dependent (coupling) rates. In general,
these rate functions may take both positive and negative values. The problem
is to characterize which constraints must be fulfilled by them in order to
obtain physically valid solutions. Interestingly, the resolution of this
issue leads us to consider all possible multipartite interaction terms, that
is, decoherence channels that involve coupling between an arbitrary number
of qubits. We also explore which rates emerge when the memory effects arise
from different underlying mechanisms based on coupling with incoherent
degrees of freedom \cite{megier,maximal}. The explicit formulation of an
operational (measurement based) memory witness \cite{modi,budiniCPF,BIF}
further provides an alternative characterization of non-Markovian effects.

As a specific example we study a family of ``hyperbolic'' and
``trigonometric'' eternal multipartite non-Markovian master equations where
some rates are negative or develop divergences at all times, respectively.
These cases provide a non-trivial extension and generalization of previous
results valid for single systems \cite{canonicalCresser}.

The paper is structured as follows. In Sec.~II we present the general class
of multipartite dynamics we consider, characterizing solution of the master
equation, resolving in consequence the constraints that guarantee the CP
condition of the map. General properties are derived for this class of
models. In Sec.~III the eternal multipartite dynamics are characterized. In
Sec.~IV we study memory effects through an operational memory witness. In
Sec.~V we provide our Conclusions. The Appendixes give details of
derivations and also obtain the rates associated to different underlying
memory mechanisms.

\section{Multipartite dynamics}

The system of interest consists of an arbitrary number $N$ of qubits. For
notational convenience we define a set of Pauli strings $S_{\mathbf{a}%
}\equiv \sigma _{a_{1}}\otimes \sigma _{a_{2}}\otimes \sigma _{a_{N}},$ each
one associated to the vector $\mathbf{a}=(a_{1},a_{2},\cdots ,a_{N}).$ Each
component $a_{k}$ $(k=1,2,\cdots N)$ assumes the values $a_{k}=(0,1,2,3)%
\leftrightarrow (\mathrm{I},\sigma _{x},\sigma _{y},\sigma _{z}),$ each one
being associated to the (two-dimensional) identity matrix and the standard
three Pauli matrices.

The evolution of the system's density matrix $\rho _{t}$\ is written in a
local-in-time way. Arbitrary multipartite decoherence channels are
considered, 
\begin{equation}
\frac{d}{dt}\rho _{t}=\mathcal{L}[\rho _{t}]=\sum_{ \mathbf{a\neq 0}}\gamma
_{t}^{\mathbf{a}}(S_{\mathbf{a}}\rho _{t}S_{\mathbf{a}}-\rho _{t}).
\label{class}
\end{equation}%
The set of functions $\{\gamma _{t}^{\mathbf{a}}\}$ define the rates
associated to the multipartite Pauli channel. In general, there are $4^{N}-1 
$ different rate functions, as the vector $\mathbf{0}=(0,0,\cdots ,0)$ is
associated to the identity operator in the full Hilbert space. Our goal is
to characterize the different aspects of this general evolution. A
time-convoluted formulation of the above dynamics is provided in Appendix~A.

\subsection{Subsystem dynamics}

Given the evolution above, we ask about the dynamics of any particular
subsystem. Introducing the splitting $\mathbf{a}=(\mathbf{a}_{\mathbf{s}},%
\mathbf{a}_{\mathbf{e}}),$ where $\mathbf{a}_{\mathbf{s}}$ corresponds to
the set of local operators that define the marginal Pauli string of the
subsystem of interest, and $\mathbf{a}_{\mathbf{e}}$ that of the rest of
qubits (now considered as part of the environment),\ from Eq.~(\ref{class})
the subsystem density matrix $\rho _{t}^{\mathbf{s}}=\mathrm{Tr}_{\mathbf{e}%
}[\rho _{t}]$ (where $\mathrm{Tr}[\bullet ]$ is the trace operation) reads%
\begin{equation}
\frac{d}{dt}\rho _{t}^{\mathbf{s}}=\sum_{\mathbf{a}_{\mathbf{s}}}\gamma
_{t}^{\mathbf{a}_{\mathbf{s}}}(S_{\mathbf{a}_{\mathbf{s}}}\rho _{t}^{\mathbf{%
s}}S_{\mathbf{a}_{\mathbf{s}}}-\rho _{t}^{\mathbf{s}}),\ \ \ \ \ \ \gamma
_{t}^{\mathbf{a}_{\mathbf{s}}}\equiv \sum_{\mathbf{a}_{\mathbf{e}}}\gamma
_{t}^{\mathbf{a}_{\mathbf{s}},\mathbf{a}_{\mathbf{e}}}.
\end{equation}%
From this equation we conclude that any subsystem, even when in general is
correlated with the complementary part, has an independent self-evolution.
In addition, the structure of this evolution belongs to the same class as
that of the full system [Eq.~(\ref{class})]. Consequently, the following
results can be particularized for any subsystem of arbitrary size.

\subsection{Solution map and completely positive condition}

We now show that by using the method of damping bases or spectral
decomposition \cite{eigen}, the solution map $\rho _{0}\rightarrow \rho _{t}$
corresponding to Eq.~(\ref{class}) can be obtained in an exact way. In order
the apply this technique, first we establish a set of relations fulfilled by
the (two dimensional) Pauli operators. Maintaining the notation $(\sigma
_{0},\sigma _{1},\sigma _{2},\sigma _{3})\leftrightarrow (\mathrm{I},\sigma
_{x},\sigma _{y},\sigma _{z}),$ it is easy to check that%
\begin{equation}
\sigma _{a}\mathrm{Tr}[\sigma _{a}\bullet ]=\frac{1}{2}\sum_{b}H_{ab}(\sigma
_{b}\bullet \sigma _{b}),  \label{HadaEigen}
\end{equation}%
where the input $[\bullet ]$ is an arbitrary two dimensional operator and $%
b=0,1,2,3.$ The inverse relation reads%
\begin{equation}
\sigma _{a}\bullet \sigma _{a}=\frac{1}{2}\sum_{b}H_{ab}\ \sigma _{b}\mathrm{%
Tr}[\sigma _{b}\bullet ].  \label{HadaKraus}
\end{equation}%
In these expressions, the coefficients $\{H_{ab}\}$ define a four
dimensional Hadamard matrix $H,$ which reads%
\begin{equation}
H\equiv \left( 
\begin{array}{cccc}
1 & 1 & 1 & 1 \\ 
1 & 1 & -1 & -1 \\ 
1 & -1 & 1 & -1 \\ 
1 & -1 & -1 & 1%
\end{array}%
\right) .  \label{Hadamard}
\end{equation}%
In deriving Eq.~(\ref{HadaKraus}), we used that its inverse reads $%
H^{-1}=H/4.$ Also notice that $H=H^T.$

Now, we introduce an extra rate $\gamma _{t}^{\mathbf{0}},$ which is
associated to the identity string in the full Hilbert space 
\begin{equation}
\gamma _{t}^{\mathbf{0}}\equiv -\sum_{\mathbf{a\neq 0}}\gamma _{t}^{\mathbf{a%
}}.  \label{GamaCero}
\end{equation}%
With this definition, the Lindbladian-like structure of Eq.~(\ref{class})]
can straightforwardly be written as 
\begin{equation}
\mathcal{L}[\bullet ]=\sum_{\mathbf{a}}\gamma _{t}^{\mathbf{a}}(S_{\mathbf{a}%
}\bullet S_{\mathbf{a}}),
\end{equation}
where the sum now includes the (identity) string $\mathbf{a=0}$. Written in
this way, applying the ``vectorial extension'' of Eq.~(\ref{HadaKraus}) to
the Hilbert space of $N$ qubits, it follows that 
\begin{equation}
\mathcal{L}[\bullet ]=\frac{1}{2^{N}}\sum_{\mathbf{a}} S_{\mathbf{a}}\mathrm{%
Tr}[ S_{\mathbf{a}}\bullet ]\sum_{\mathbf{b}}H_{\mathbf{ab}}\gamma _{t}^{%
\mathbf{b}},
\end{equation}%
where $H_{\mathbf{ab}}\equiv H_{a_{1}b_{1}}H_{a_{2}b_{2}}\cdots
H_{a_{N}b_{N}}$ can be read as the matrix elements of the external product
of $N$ single Hadamard matrices, cf.\ Eq. (\ref{Hadamard}). From this last
expression, by using that $\mathrm{Tr}[S_{\mathbf{a}}S_{\mathbf{b}%
}]=2^{N}\delta _{\mathbf{a},\mathbf{b}},$ it is straightforward to determine
the eigenvalues and eigenoperators of $\mathcal{L}[\bullet ].$\ They read%
\begin{equation}
\mathcal{L}[S_{\mathbf{a}}]=\mu _{t}^{\mathbf{a}} S_{\mathbf{a}},\ \ \ \ \ \
\ \ \ \ \mu _{t}^{\mathbf{a}}=\sum_{\mathbf{b}}H_{\mathbf{ab}}\gamma _{t}^{%
\mathbf{b}}.  \label{Eigen}
\end{equation}%
Consequently, any Pauli string $S_{\mathbf{a}}$ is a right eigenoperator
with eigenvalue $\mu _{t}^{\mathbf{a}}.$\ Given that $\mathcal{L}[\bullet ]$
also defines the adjoint evolution (as the ``jump operators'' are Hermitian) 
\cite{eigen}, $S_{\mathbf{a}}$ is also a left eigenoperator. Notice also
that by using the inverse of the Hadamard matrix, the inverse relation $%
\gamma _{t}^{\mathbf{a}}=\sum_{\mathbf{b}}H_{\mathbf{ab}}\mu _{t}^{\mathbf{b}%
}/4^{N}$ follows.

From the method of damping bases \cite{eigen}, Eq.~(\ref{Eigen}) allows us
to write the solution of Eq. (\ref{class}) as%
\begin{equation}
\rho _{t}=\frac{1}{2^{N}}\sum_{\mathbf{a}}\exp \left[ \int_{0}^{t}dt^{\prime
}\mu _{t^{\prime }}^{\mathbf{a}}\right] S_{\mathbf{a}}\mathrm{Tr}[S_{\mathbf{%
a}}\rho _{0}].  \label{SolutionEigen}
\end{equation}%
One can see that the conditions $\mathrm{Tr}[\rho _{t}]=\mathrm{Tr}[\rho
_{0}]=1$ are satisfied after noting that $\mathrm{Tr}[S_{\mathbf{a}%
}]=2^{N}\delta _{\mathbf{a},\mathbf{0}}$ and $\mu _{t^{\prime }}^{\mathbf{0}%
}=0.$ This last equality follows from Eqs.~(\ref{GamaCero}) and (\ref{Eigen}%
) jointly with the property $H_{\mathbf{0b}}=1$ $\forall \mathbf{b.}$ By
using the vectorial extension of Eq.~(\ref{HadaEigen}), we get the density
matrix written in a Kraus representation \cite{alicki,breuerbook},%
\begin{equation}
\rho _{t}=\sum_{\mathbf{a}}p_{t}^{\mathbf{a}}(S_{\mathbf{a}}\rho _{0}S_{%
\mathbf{a}}).  \label{Solution}
\end{equation}%
The weights are $p_{t}^{\mathbf{a}}=4^{-N}\sum_{\mathbf{b}}H_{\mathbf{ab}%
}\exp [\int_{0}^{t}dt^{\prime }\mu _{t^{\prime }}^{\mathbf{b}}],$ which from
Eq. (\ref{Eigen}) can explicitly be written in terms of the time-dependent
rates as%
\begin{equation}
p_{t}^{\mathbf{a}}=\frac{1}{4^{N}}\sum_{\mathbf{b}}H_{\mathbf{ab}}\exp \left[
\sum_{\mathbf{c}}H_{\mathbf{bc}}\int_{0}^{t}dt^{\prime }\gamma _{t^{\prime
}}^{\mathbf{c}}\right] .  \label{ProbSol}
\end{equation}

The final expressions (\ref{Solution}) and (\ref{ProbSol}) are the main
results of this section. They completely characterize the solution map in
terms of the set of rates $\{\gamma _{t}^{\mathbf{a}}\}$ and the initial
condition $\rho _{0}.$ In addition, they naturally provide a constraint that
the rates must to fulfill in order to obtain a CP map, that is, one that
gives physical solution. In fact, the Kraus representation theorem \cite%
{alicki,breuerbook} implies the conditions $0\leq p_{t}^{\mathbf{a}}\leq 1,$
which means that $\{p_{t}^{\mathbf{a}}\}$\ are a set of normalized
probabilities. In the single qubit case $(N=1),$ previously obtained
constraints are recovered \cite{wudarski}. In the general case, $4^{N}$
inequalities must be fulfilled. We notice that a sufficient, but not
necessary, condition is $\int_{0}^{t}dt^{\prime }\gamma _{t^{\prime }}^{%
\mathbf{a}}\geq 0$ $\forall \mathbf{a\neq 0.}$ In fact, this constraint
implies that all eigenvalues, cf.\ Eq.~(\ref{Eigen}), satisfy $\mu _{t}^{%
\mathbf{a}}\leq 0$ $(\mathbf{a}\neq \mathbf{0}).$ Consequently, taking an
arbitrary but \textit{fixed} time $t,$ the solution (\ref{SolutionEigen}) of
the non-Markovian dynamics, via the association $\int_{0}^{t}dt^{\prime }\mu
_{t^{\prime }}^{\mathbf{a}}=t\mu _{M}^{\mathbf{a}},$ is equivalent to the
solution of a (well behaved) Markovian dynamics generated by a Lindbladian
with eigenvalues $\{\mu _{M}^{\mathbf{a}}\}.$

\subsection{Non-Markovianity and time-dependent rates}

Different (inequivalent) memory witnesses based only on the system
propagator can be used to define non-Markovianity \cite%
{BreuerReview,plenioReview} such as for example the trace distance between
two different initial conditions \cite{BreuerFirst} or those based on the $k$%
-positivity of the solution map \cite{DarioSabrina}. Here, as the dynamics
is written naturally in a canonical form \cite{canonicalCresser}, memory
effects can also be defined by the negativity of the time-dependent rates $%
\{\gamma _{t}^{\mathbf{a}}\}.$ In this way, it is of interest to determine
these elements for any well behaved solution defined by the probabilities $%
\{p_{t}^{\mathbf{a}}\}$ in Eq. (\ref{Solution}).

We can invert Eq.~(\ref{ProbSol}), 
\begin{equation}
\mu _{t}^{\mathbf{a}}=\frac{d}{dt}\ln \left[ \sum_{\mathbf{b}}H_{\mathbf{ab}%
}p_{t}^{\mathbf{b}}\right] ,
\end{equation}%
and using Eq.~(\ref{Eigen}) we get explicit expressions for the set of rates 
$\{\gamma _{t}^{\mathbf{a}}\}$ in terms of the normalized time-dependent
weights $0\leq p_{t}^{\mathbf{c}}\leq 1$, 
\begin{equation}
\gamma _{t}^{\mathbf{a}}=\frac{1}{4^{N}}\sum_{\mathbf{b}}H_{\mathbf{ab}}%
\frac{d}{dt}\ln \left[ \sum_{\mathbf{c}}H_{\mathbf{bc}}p_{t}^{\mathbf{c}}%
\right] .  \label{SolRates}
\end{equation}%
The signs of $\{\gamma _{t}^{\mathbf{a}}\}$ can be taken as a signature of
departure from a Markovian regime \cite{canonicalCresser}. Alternatively, in
Sec.~V we study operational measures for non-Markovianity. We notice that
Eqs.~(\ref{ProbSol}) and~(\ref{SolRates}) provide a multipartite
generalization of the case $N=1$ studied in Ref.~\cite{wudarski}.

\subsection{Additivity of non-Markovian master equations}

Given two sets of (arbitrary) normalized probabilities $\{p_{t}^{\mathbf{a}%
}\}$ and $\{\tilde{p}_{t}^{\mathbf{a}}\},$ the relation (\ref{SolRates})
allows us to obtain the corresponding sets of rates $\{\gamma _{t}^{\mathbf{a%
}}\}$ and $\{\tilde{\gamma}_{t}^{\mathbf{a}}\}.$ From these we can obtain a
new master equation defined by Eq.~(\ref{class}) with rates $\{\gamma _{t}^{%
\mathbf{a}}+\tilde{\gamma}_{t}^{\mathbf{a}}\}$. In fact, it is always
possible to associate a set of probabilities $\{q_{t}^{\mathbf{a}}\}$ to
these added rates, that is,%
\begin{equation}
\{p_{t}^{\mathbf{a}}\}\leftrightarrow \{\gamma _{t}^{\mathbf{a}}\},\ \ \ \{%
\tilde{p}_{t}^{\mathbf{a}}\}\leftrightarrow \{\tilde{\gamma}_{t}^{\mathbf{a}%
}\},\ \ \Rightarrow \ \ \exists \{q_{t}^{\mathbf{a}}\}\leftrightarrow
\{\gamma _{t}^{\mathbf{a}}+\tilde{\gamma}_{t}^{\mathbf{a}}\}.
\label{Additivity}
\end{equation}%
Consequently, as occurs to Markovian Lindblad equations~\cite{breuerbook},
for our class of models arbitrary well behaved evolutions (defined by a
given set of rates) can be added in an arbitrary way. The validity of this
result follows from the commutation of two arbitrary propagators, Eq.~(\ref%
{Solution}), a property supported by the relation%
\begin{equation}
S_{\mathbf{a}}S_{\mathbf{b}}\bullet S_{\mathbf{b}}S_{\mathbf{a}}=S_{\mathbf{b%
}}S_{\mathbf{a}}\bullet S_{\mathbf{a}}S_{\mathbf{b}}=S_{\mathbf{c}}\bullet
S_{\mathbf{c}}^{\dag },  \label{SanguchesPaulis}
\end{equation}%
which is valid for arbitrary Pauli strings $S_\mathbf{a}$ and $S_\mathbf{b},$
and where $S_{\mathbf{c}}=S_{\mathbf{a}}S_{\mathbf{b}}$ or equivalently $S_{%
\mathbf{c}}=S_{\mathbf{b}}S_{\mathbf{a}}.$ Eq.~(\ref{SanguchesPaulis}) can
be straightforwardly demonstrated from Eq.~(\ref{HadaKraus}).

\subsection{Coupling with incoherent degrees of freedom}

Memory effects are induced whenever extra degrees of freedom are traced out.
Here, we consider a general coupling with incoherent degrees of freedom.
Based on Ref.~\cite{lindbladrate}, the more general case can always be
described by writing the system density matrix $\rho _{t}$ and the
probabilities of the incoherent system $\{q_{t}^{\mathbf{h}}\}$ as%
\begin{equation}
\rho _{t}=\sum_{\mathbf{h}}\rho _{t}^{\mathbf{h}},\ \ \ \ \ \ \ \ \ \ \
q_{t}^{\mathbf{h}}=\mathrm{Tr}[\rho _{t}^{\mathbf{h}}],
\label{IncoherentStructure}
\end{equation}%
where the auxiliary states $\{\rho _{t}^{\mathbf{h}}\}$ correspond to the
system state given that the extra (hidden) incoherent degrees of freedom are
in the particular state $\mathbf{h.}$ The evolution of the states $\{\rho
_{t}^{\mathbf{h}}\}$ may involve coupling between all of them \cite%
{lindbladrate}. Given the structure Eq.~(\ref{class}), each auxiliary state $%
\rho _{t}^{\mathbf{h}}$ must to assume the form%
\begin{equation}
\rho _{t}^{\mathbf{h}}=\sum_{\mathbf{\alpha }}g_{\mathbf{\alpha }}^{\mathbf{h%
}}(t)(S_{\mathbf{\alpha }}\rho _{0}S_{\mathbf{\alpha }}),  \label{RhoAux}
\end{equation}%
where the parameter $\mathbf{\alpha }$ runs over a set of Pauli strings that
depends on each specific problem. The functions $g_{\mathbf{\alpha }}^{%
\mathbf{h}}(t)$ in turn obey a classical master equation whose structure
also depends on each specific model. The initial conditions read $\rho _{0}^{%
\mathbf{h}}=\rho _{0}q_{0}^{\mathbf{h}},$ where $\rho _{0}$\ is the initial
system state and $q_{0}^{\mathbf{h}}$\ is the initial probability of the
incoherent degrees of freedom. In fact, at time $t,$ $q_{t}^{\mathbf{h}%
}=\sum_{\mathbf{\alpha }}g_{\mathbf{\alpha }}^{\mathbf{h}}(t).$ On the other
hand, the system density matrix evolution [Eq.~(\ref{Solution})] is defined
by the probabilities $p_{t}^{\mathbf{\alpha }}=\sum_{\mathbf{h}}g_{\mathbf{%
\alpha }}^{\mathbf{h}}(t).$ A general treatment is not possible. Relevant
examples are worked out in Appendix B such as a mapping with a classical
Markovian master equation, stochastic Hamiltonians, and statistical mixtures
of Markovian evolutions. In all cases, explicit expressions for the rates
[Eq.~(\ref{SolRates})] can be obtained. A representative class of dynamics
is studied in the next section.

\section{Multipartite eternal non-Markovianity}

For a single qubit, $N=1$, the system density matrix evolution, Eq.~(\ref%
{class}), may involve rates that are negative at all times. This property
was called ``eternal non-Markovianity'' \cite{canonicalCresser,megier}. The
results of Appendix~B [see Eqs.~(\ref{GamaEternalMaster}), (\ref{RateHst}),
and (\ref{GamaTwoMixture})] and Appendix~C [see Eqs.~(\ref{GamaBip}) and (%
\ref{GamaTrip})] guarantee that this property also emerges in multipartite
dynamics, $N>1$, which have $4^{N}-1$ rates.

In order to provide simple (multipartite) examples, here we restrict\ to the
case where the evolution is 
\begin{eqnarray}
\mathcal{L}[\bullet ] &=&\Big{\{}\gamma _{t}^{\underline{\mathbf{a}}}(S_{%
\underline{\mathbf{a}}}\bullet S_{\underline{\mathbf{a}}}-\bullet )+\gamma
_{t}^{\underline{\mathbf{b}}}(S_{\underline{\mathbf{b}}}\bullet S_{%
\underline{\mathbf{b}}}-\bullet )  \notag \\
&&+\gamma _{t}^{\underline{\mathbf{c}}}(S_{\underline{\mathbf{c}}}\bullet S_{%
\underline{\mathbf{c}}}^{\dagger }-\bullet )\Big{\}},
\end{eqnarray}%
where $S_{\underline{\mathbf{a}}}\mathbf{\ }$and\textbf{\ }$S_{\underline{%
\mathbf{b}}}$ are two arbitrary multipartite Pauli strings, while $S_{%
\underline{\mathbf{c}}}=S_{\underline{\mathbf{a}}}S_{\underline{\mathbf{b}}%
}. $ Depending on the time-dependence of the rates we define what we term
\textquotedblleft hyperbolic\textquotedblright\ and \textquotedblleft
trigonometric\textquotedblright\ cases of eternal non-Markovianity.

\subsection{Hyperbolic eternal non-Markovianity}

The system density matrix is written as the addition of two auxiliary
states\ $\rho _{t}=\rho _{t}^{(1)}+\rho _{t}^{(2)}$ [Eq.~(\ref%
{IncoherentStructure})], whose evolution reads 
\begin{subequations}
\label{RhoEternalViaSuperposition}
\begin{eqnarray}
\frac{d\rho _{t}^{(1)}}{dt} &=&-\gamma \rho _{t}^{(1)}+\gamma S_{\underline{%
\mathbf{a}}}\rho _{t}^{(1)}S_{\underline{\mathbf{a}}}, \\
\frac{d\rho _{t}^{(2)}}{dt} &=&-\varphi \rho _{t}^{(2)}+\varphi S_{%
\underline{\mathbf{b}}}\rho _{t}^{(2)}S_{\underline{\mathbf{b}}}.
\end{eqnarray}%
The initial conditions for the auxiliary states are taken to be $\rho
_{0}^{(1)}=\rho _{0}^{(2)}=\rho_{0}/2.$ Given that the auxiliary states do
not couples, the rates of the non-Markovian evolution follow Eq.~(\ref%
{SolRates}) with probabilities $p_{t}^{\mathbf{a}}=[p_{1}^{\mathbf{a}%
}(t)+p_{2}^{\mathbf{a}}(t)]/2$, with the sets $\{p_{1}^{\mathbf{a}}(t)\}$
and $\{p_{2}^{\mathbf{a}}(t)\},$ via Eq. (\ref{ProbSol}), associated to $%
\rho _{t}^{(1)}$ and $\rho _{t}^{(2)}$, respectively. Taking $\varphi
=\gamma ,$ we get [see also derivation from Eq.~(\ref{GamaTwoMixture}) in
Appendix C] 
\end{subequations}
\begin{equation}
\gamma _{t}^{\underline{\mathbf{a}}}=\gamma _{t}^{\underline{\mathbf{b}}}=%
\frac{1}{2}\gamma ,\ \ \ \ \ \ \ \gamma _{t}^{\underline{\mathbf{c}}}=-\frac{%
1}{2}\gamma \tanh (\gamma t).  \label{Tanh}
\end{equation}%
This result provides a multipartite generalization, $(N>1)$, of the single
qubit case $(N=1)$ studied in Ref.~\cite{canonicalCresser}. Similarly to the
results of Ref.~\cite{megier} we notice that in this particular case
alternative dynamics such as the mapping to a classical master equation [see
Eq.~(\ref{GamaEternalMaster})] and stochastic Hamiltonians [see Eq.~(\ref%
{RateHst})] also lead to the same rates.

\subsection{Trigonometric eternal non-Markovianity}

Based on Eq.~(\ref{IncoherentStructure}), instead of the evolution (\ref%
{RhoEternalViaSuperposition}), here we consider 
\begin{subequations}
\label{RhoTrigonometric}
\begin{eqnarray}
\frac{d\rho _{t}^{(1)}}{dt} &=&-\gamma \rho _{t}^{(1)}+\varphi S_{\underline{%
\mathbf{b}}}\rho _{t}^{(2)}S_{\underline{\mathbf{b}}}, \\
\frac{d\rho _{t}^{(2)}}{dt} &=&-\varphi \rho _{t}^{(2)}+\gamma S_{\underline{%
\mathbf{a}}}\rho _{t}^{(1)}S_{\underline{\mathbf{a}}}.
\end{eqnarray}%
The initial conditions are taken as $\rho _{0}^{(1)}=[\varphi /(\varphi
+\gamma )]\rho _{0}$ and $\rho _{0}^{(2)}=[\gamma /(\varphi +\gamma )]\rho
_{0},$ where $\rho _{0}$ is the system initial state. Notice that the
incoherent transitions $(1)\leftrightarrow (2)$ imply the system
transformations $\rho \rightarrow S_{\underline{\mathbf{a}}/\underline{%
\mathbf{b}}} \rho S_{\underline{\mathbf{a}}/\underline{\mathbf{b}}}$.

Taking into account Eq.~(\ref{RhoAux}), in order to solve Eq.~(\ref%
{RhoTrigonometric}) each auxiliary state is written as $(h=1,2)$%
\end{subequations}
\begin{equation}
\rho _{t}^{(h)}=g_{\mathbf{0}}^{(h)}\rho _{0}+g_{\underline{\mathbf{a}}%
}^{(h)}S_{\underline{\mathbf{a}}}\rho _{0}S_{\underline{\mathbf{a}}}+g_{%
\underline{\mathbf{b}}}^{(h)}S_{\underline{\mathbf{b}}}\rho _{0}S_{%
\underline{\mathbf{b}}}+g_{\underline{\mathbf{c}}}^{(h)} S_{\underline{%
\mathbf{c}}}\rho _{0}S_{\underline{\mathbf{c}}}^{\dagger },
\label{AlfaBetaRun}
\end{equation}%
where as before $S_{\underline{\mathbf{c}}}=S_{\underline{\mathbf{a}}}S_{%
\underline{\mathbf{b}}}$, and $g_{\mathbf{\alpha }}^{(h)}$ are
time-dependent functions. Using Eq.~(\ref{SanguchesPaulis}), it is possible
to derive a classical master equation for the (eight) $g$-functions which
involves coupling between pairs of them. The corresponding solutions allow
to obtain the probabilities $p_{t}^{\mathbf{a}}=\sum_{h}g_{\mathbf{a}%
}^{(h)}(t).$ Finally, the rates associated to the non-Markovian evolution
follow from Eq.~(\ref{SolRates}) 
\begin{equation}
\gamma _{t}^{\underline{\mathbf{a}}}=\gamma _{t}^{\underline{\mathbf{b}}}=%
\frac{\varphi \gamma (\varphi +\gamma )}{e^{t(\varphi +\gamma )}(\varphi
-\gamma )^{2}+4\varphi \gamma }.  \label{GamaATrig}
\end{equation}%
Furthermore,%
\begin{eqnarray}
\gamma _{t}^{\underline{\mathbf{c}}} &=&\varphi \gamma \{\delta _{+}\Upsilon
^{2}(1-e^{t\Upsilon })-\delta _{-}^{2}[\Upsilon (1+e^{t\Upsilon })  \notag \\
&&+e^{t\delta _{+}}[(\delta _{+}-\Upsilon )-e^{t\Upsilon }(\delta
_{+}+\Upsilon )]]\}  \notag \\
&&\times \{(e^{t\delta _{+}}\delta _{-}^{2}+4\varphi \gamma
)[(1+e^{t\Upsilon })\Upsilon \delta _{+}  \notag \\
&&-(1-e^{t\Upsilon })\delta _{-}^{2}]\}^{-1},  \label{GamaCTrig}
\end{eqnarray}%
where the coefficients are%
\begin{equation}
\Upsilon \equiv (\varphi ^{2}-6\varphi \gamma +\gamma ^{2})^{1/2},\ \ \ \ \
\ \ \delta _{\pm }\equiv \varphi \pm \gamma .  \label{ParameterTransition}
\end{equation}

Depending on the ratio $\varphi /\gamma ,$ different characteristic
behaviors are obtained. In Fig.~1 we plot both rates. Consistent with Eq.~(%
\ref{GamaATrig}), $\gamma _{t}^{\underline{\mathbf{a}}}$ and $\gamma _{t}^{%
\underline{\mathbf{b}}}$ are always positive functions. However, this is not
the case for $\gamma _{t}^{\underline{\mathbf{c}}},$ Eq.~(\ref{GamaCTrig}),
which depending on $\varphi /\gamma $ develops a transition between \textit{%
positivity} [Figs.~5(a) and 5(d)] and a \textit{periodic divergent behavior}
[Figs.~5(b) and 5(c)]. From Eq.~(\ref{ParameterTransition}) we deduce that
this change occurs in the boundaries of the interval $3-\sqrt{8}<(\varphi
/\gamma )<3+\sqrt{8}$, with $\gamma _{t}^{\underline{\mathbf{c}}}$
developing divergences in this interval, while being positive outside it.

From the plots it is also evident that $\gamma _{t}^{\underline{\mathbf{a}}}$
and $\gamma _{t}^{\underline{\mathbf{b}}}$ approach a constant when $\varphi
\approx \gamma .$ In fact, when $\varphi =\gamma ,$ the previous expressions
reduce to%
\begin{equation}
\gamma _{t}^{\underline{\mathbf{a}}}=\gamma _{t}^{\underline{\mathbf{b}}}=%
\frac{1}{2}\gamma ,\ \ \ \ \ \ \ \gamma _{t}^{\underline{\mathbf{c}}}=\frac{1%
}{2}\gamma \tan (\gamma t).  \label{TanTrig}
\end{equation}%
Based on Eq.~(\ref{Tanh}), we name this case as a \emph{trigonometric
eternal non-Markovian}. The probabilities $\{p_{t}^{\mathbf{a}}\}$ [Eq.~(\ref%
{Solution})] also assume a simple form, 
\begin{subequations}
\begin{eqnarray}
p_{t}^{\mathbf{0}} &=&\frac{1}{2}e^{-\gamma t}[\cosh (\gamma t)+\cos (\gamma
t)], \\
p_{t}^{\underline{\mathbf{a}}} &=&p_{t}^{\underline{\mathbf{b}}}=\frac{1}{4}%
[1-e^{-2\gamma t}], \\
p_{t}^{\underline{\mathbf{c}}} &=&\frac{1}{2}e^{-\gamma t}[\cosh (\gamma
t)-\cos (\gamma t)].
\end{eqnarray}%
These solutions apply to arbitrary multipartite Pauli strings $\underline{%
\mathbf{a}}$ and $\underline{\mathbf{b}}.$%
%figura1%figura%figura%figura%figurav%figura%figura%figura%figura%figura%figura%figura%figura%figura%figurav%figura%figura%figura%figura%figura
%figura%figura%figura%figura%figurav%figura%figura%figura%figura%figura%figura%figura%figura%figura%figurav%figura%figura%figura%figura%figura
\begin{figure}[tbp]
\includegraphics[bb=38 598 725
1140,angle=0,width=8.5cm]{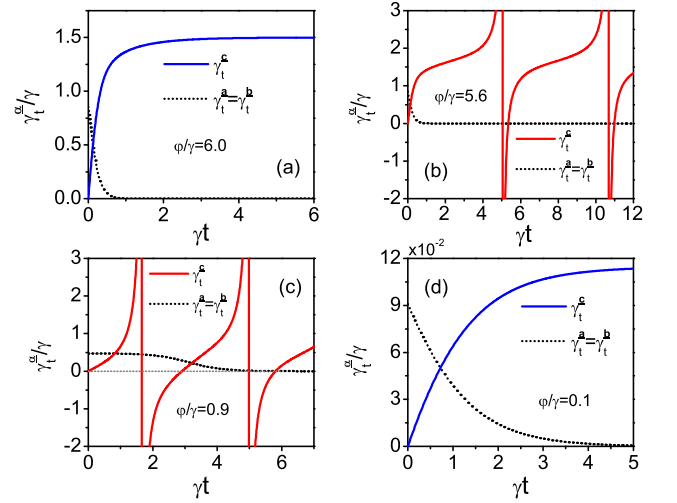} 
\caption{Time dependent rates [Eqs.~(\protect\ref{GamaATrig}) and (\protect
\ref{GamaCTrig})] corresponding to the multipartite trigonometric eternal
non-Markovian evolution [Eq.~(\protect\ref{RhoTrigonometric})] for different
values of the rate ratio $\protect\varphi /\protect\gamma .$}
\end{figure}
%figura%figura%figura%figura%figurav%figura%figura%figura%figura%figura%figura%figura%figura%figura%figurav%figura%figura%figura%figura%figura
%figura%figura%figura%figura%figurav%figura%figura%figura%figura%figura%figura%figura%figura%figura%figurav%figura%figura%figura%figura%figura

\subsection{Adding non-Markovian evolutions}

Added to the previous examples (see also Appendix C), the possibility of
adding arbitrary (well defined) rates [Eq.~(\ref{Additivity})] gives us a
procedure for constructing a large family of well behaved dynamics. For
example, we write 
\end{subequations}
\begin{eqnarray}
\mathcal{L}[\bullet ] &=&\sum_{i=1}^{N}\frac{\gamma _{i}}{2}\Big{\{}(\sigma
_{i}^{x}\sigma _{i+1}^{x}\bullet \sigma _{i+1}^{x}\sigma _{i}^{x}-\bullet ) 
\notag \\
&&\ \ \ \ \ \ \ +(\sigma _{i}^{y}\sigma _{i+1}^{y}\bullet \sigma
_{i+1}^{y}\sigma _{i}^{y}-\bullet )  \label{traslacional} \\
&&\ \ \ \ \ \ \ +f_{i}(t)(\sigma _{i}^{z}\sigma _{i+1}^{z}\bullet \sigma
_{i+1}^{z}\sigma _{i}^{z}-\bullet )\Big{\}}.  \notag
\end{eqnarray}%
In this traslational invariant generator (say with periodic boundaries in
one dimension), we may chose $f_{i}(t)=-\tanh (\gamma _{i}t)$ or
alternatively $f_{i}(t)=\tan (\gamma _{i}t)$ [see Eqs.~(\ref{Tanh}) and (\ref%
{TanTrig}) respectively].

One interesting aspect of using additivity for constructing multipartite
evolutions is that, even when the underlying evolutions have a clear memory
mechanism (see also Appendix B), the resulting dynamics does not
necessarily. For example, while our approach guarantees that Eq.~(\ref%
{traslacional}) leads to a completely positive dynamics [with solution
defined by Eqs.~(\ref{Solution}) and (\ref{ProbSol})] it is not evident
which underlying processes may lead to this master equation. In addition, in
general there may be subsystems that are coupled between then, one part
being Markovian and the other non-Markovian. For example, take $%
f_{i}(t)=\gamma _{i}/2$ for $i\leq N_{0},$ and $f_{i}(t)=\tan (\gamma _{i}t)$
for $i>N_{0}.$

\section{Operational memory witness}

An alternative and deeper characterization of quantum non-Markovianity can
be obtained by defining memory effects via measurement based approaches~\cite%
{modi,budiniCPF,BIF}. Here, we study a conditional past-future (CPF)
correlation~\cite{budiniCPF}. This object relies on performing three
successive measurement of arbitrary system observables and calculating the
correlation between the last (future) and first (past) outcomes conditioned
to a given intermediate (present) outcome. For Markovian dynamics it
vanishes identically, while memory effects leads to a non null CPF
correlation.

The measurements, denoted in successive order by $\underline{\mathbf{x}},$ $%
\underline{\mathbf{y}},$ and $\underline{\mathbf{z}},$ correspond to
observations of three Hermitian operators $S_{\underline{\mathbf{m}}}$ with
eigenvectors~$\{|m\rangle \}$ and eigenvalues~$\{m\},$ 
\begin{equation}
S_{\underline{\mathbf{m}}}|m\rangle =m|m\rangle ,\ \ \ \ \ \underline{%
\mathbf{m}}=\underline{\mathbf{x}},\underline{\mathbf{y}},\underline{\mathbf{%
z}}.  \label{Observables}
\end{equation}%
The CPF correlation then reads~\cite{budiniCPF}%
\begin{equation}
C_{pf}(t,\tau )|_{y}=\sum_{z,x}zx[P(z,x|y)-P(z|y)P(x|y)],
\label{CPFexplicit}
\end{equation}%
where $\{x\},$ $\{y\},$ and $\{z\}$ denotes the three sets of successive
outcomes (operators eigenvalues), while $t$ and $\tau $\ are the (first and
second) time intervals between the successive measurements. With $P(u|v)$ we
denote the conditional probability of $u$ given $v.$

All probabilities appearing in Eq.~(\ref{CPFexplicit}) can be determine from
the (outcomes) joint probability $P(z,y,x)\leftrightarrow P(z,t+\tau
,y,t;x,0),$ which in turn can be calculated after knowing the underlying
system-environment dynamics. In Appendix D we show that $P(z,y,x)$ and\ $%
C_{pf}(t,\tau )|_{y}$ can be calculated exactly assuming that memory effects
emerge due to the coupling with incoherent degrees of freedom [Eqs.~(\ref%
{IncoherentStructure}) and (\ref{RhoAux})].

Each specific model [see examples (\ref{RhoEternalViaSuperposition}) and (%
\ref{RhoTrigonometric})] is completely defined by the set of functions $\{g_{%
\mathbf{\alpha }}^{\mathbf{h}}(t)\}$ [Eq.~(\ref{RhoAux})].\ Given that they
obey a (linear) classical master equation, they can be written as%
\begin{equation}
g_{\mathbf{\alpha }}^{\mathbf{h}}(t)=\sum_{\mathbf{h}^{\prime }}f_{\mathbf{%
\alpha }}^{\mathbf{hh}^{\prime }}(t)q_{0}^{\mathbf{h}^{\prime }}\equiv (%
\mathbf{h}|\mathbb{F}_{\mathbf{\alpha }}(t)|q_{0}),  \label{gVectorial}
\end{equation}%
where the set of functions $\{f_{\mathbf{\alpha }}^{\mathbf{hh}^{\prime
}}(t)\}$ are independent of the initial conditions $\{q_{0}^{\mathbf{h}}\}.$
Furthermore, for notational simplicity, we introduced a vectorial orthogonal
base $\{|\mathbf{h})\}$ for the incoherent degrees of freedom, such that $f_{%
\mathbf{\alpha }}^{\mathbf{hh}^{\prime }}(t)\leftrightarrow (\mathbf{h}|%
\mathbb{F}_{\mathbf{\alpha }}(t)|\mathbf{h}^{\prime })$ and $q_{0}^{\mathbf{h%
}}\leftrightarrow (\mathbf{h}|q_{0}).$

The observables $S_{\underline{\mathbf{m}}}$\ [Eq.~(\ref{Observables})] may
in principle be defined by arbitrary linear combinations of Pauli strings $%
\{S_{\mathbf{a}}\}.$ Here, for simplicity they are defined by a unique Pauli
string. In this case, the general expression for the CPF correlation [Eq.~(%
\ref{CPFGen})] reduces to (see Appendix D)%
%TCIMACRO{\TeXButton{widetext}{\begin{widetext}}}%
%BeginExpansion
\begin{widetext}%
%EndExpansion
\begin{equation}
C_{pf}(t,\tau )|_{y}=\delta _{\underline{\mathbf{z}},\mathbf{y}}\delta _{%
\underline{\mathbf{y}},\underline{\mathbf{x}}}\frac{(1-\langle x\rangle ^{2})%
}{[2^{N}P(y)]^{2}}\sum_{\mathbf{\alpha ,\beta }}H_{\underline{\mathbf{y}}%
\mathbf{\alpha }}H_{\underline{\mathbf{y}}\mathbf{\beta }}[(1|\mathbb{F}_{%
\mathbf{\alpha }}(\tau )\mathbb{F}_{\mathbf{\beta }}(t)|q_{0})-(1|\mathbb{F}%
_{\mathbf{\alpha }}(\tau )|q_{t})(1|\mathbb{F}_{\mathbf{\beta }}(t)|q_{0})].
\label{CPFSol}
\end{equation}%
%TCIMACRO{\TeXButton{widetext}{\end{widetext}} }%
%BeginExpansion
\end{widetext}
%EndExpansion
In here, $|q_{t})=\sum_{\mathbf{\alpha }}\mathbb{F}_{\mathbf{\alpha }%
}(t)|q_{0})$ define the probabilities of the incoherent degrees of freedom
at time $t,$ while $(1|\equiv \sum_{\mathbf{h}}(\mathbf{h}|.$ Furthermore, $%
\langle x\rangle \equiv \sum_{x}xP(x)$ where $P(x)=\langle x|\rho
_{0}|x\rangle .$ Finally, $P(y)$ is the probability for the outcomes of the
second measurement. It is%
\begin{equation}
P(y)=\frac{1}{2^{N}}\Big{[}1+y\langle x\rangle \delta _{\underline{\mathbf{y}%
},\underline{\mathbf{x}}}\sum_{\mathbf{\alpha }}H_{\underline{\mathbf{y}}%
\mathbf{\alpha }}(1|\mathbb{F}_{\mathbf{\alpha }}(t)|q_{0})\Big{]}.
\end{equation}

The term $\delta _{\underline{\mathbf{z}},\mathbf{y}}\delta _{\underline{%
\mathbf{y}},\underline{\mathbf{x}}}$ in Eq.~(\ref{CPFSol}) implies that,
for\ observables defined by unique Pauli strings, memory effects are
detected only when the three observables are the same $S_{\underline{\mathbf{%
x}}}=S_{\underline{\mathbf{y}}}=S_{\underline{\mathbf{z}}}.$ This constraint
does not emerge when the observables correspond to other basis of operators
(see for example Ref.~\cite{BIF}).

The general solution Eq.~(\ref{CPFSol}) can be specified for the
trigonometric eternal model [Eq.~(\ref{RhoTrigonometric})]. Stationary
initial conditions are assumed, $|q_{t})=|q_{0}),$ with $q_{0}^{(1)}=\varphi
/(\varphi +\gamma )$ and $q_{0}^{(2)}=\gamma /(\varphi +\gamma ).$ For
simplicity, first we consider the case $N=1.$ When the three measurements
are performed in direction $\mathbf{\underline{a}}$ or $\mathbf{\underline{b}%
}$ we get%
\begin{eqnarray}
C_{pf}(t,\tau )|_{y} &=&-\frac{(1-\langle x\rangle ^{2})}{[2^{N}P(y)]^{2}}%
\exp \left[ -(t+\tau )(\gamma +\varphi )/2\right]  \notag \\
&&\!\!\!\!\!\!\!\times \frac{4^{2}\gamma ^{2}\varphi ^{2}}{(\gamma +\varphi
)^{2}\Upsilon ^{2}}\sinh \left( \frac{\Upsilon t}{2}\right) \sinh \left( 
\frac{\Upsilon \tau }{2}\right) .\ \ \ \ \   \label{CPFTrigAB}
\end{eqnarray}%
When the three measurements are performed in direction $\mathbf{\underline{c}%
,}$ we get%
\begin{eqnarray}
C_{pf}(t,\tau )|_{y} &=&\frac{(1-\langle x\rangle ^{2})}{[2^{N}P(y)]^{2}}\ 
\frac{4\gamma \varphi (\gamma -\varphi )^{2}}{(\gamma +\varphi )^{4}}
\label{CPFZeta} \\
&&\times \lbrack 1-e^{-\tau (\gamma +\varphi )}][1-e^{-t(\gamma +\varphi )}].
\notag
\end{eqnarray}

These results allow us to analyze the transition to divergent rates [Eq.~(%
\ref{GamaCTrig})] in a complementary way. In Fig.~2 we plot the CPF
correlation Eq.~(\ref{CPFTrigAB}). We observe that when the rate $\gamma
_{t}^{\underline{\mathbf{c}}}$ does not develop divergences [Fig.~2(a)], the
CPF correlation is negative for any value of the time intervals $t$ and $%
\tau .$ On the other hand, in the interval $3-\sqrt{8}<(\varphi /\gamma )<3+%
\sqrt{8}$ where the rate $\gamma _{t}^{\underline{\mathbf{c}}}$ develops
divergences [Fig.~2(b)], the CPF correlation presents oscillations between
positive an negative values.

For the model~(\ref{RhoTrigonometric}), the generalization to $N>1,$
independently of the chosen observables, always lead to Eq.~(\ref{CPFTrigAB}%
) or Eq.~(\ref{CPFZeta}). This results follows by noting that in Eq.~(\ref%
{CPFSol}) the coefficients $\mathbf{\alpha }$ and $\mathbf{\beta }$ only
assume the four values $\mathbf{\alpha ,\beta }=(\mathbf{0,}\underline{%
\mathbf{a}},\underline{\mathbf{b}},\underline{\mathbf{c}})$ [see Eq.~(\ref%
{AlfaBetaRun})]. Furthermore, using that $H_{\underline{\mathbf{y}}\mathbf{%
\alpha }}H_{\underline{\mathbf{y}}\mathbf{\beta }}=H_{\underline{\mathbf{y}}%
\mathbf{\gamma }},$ where $\mathbf{\gamma }$ corresponds to the string $S_{%
\mathbf{\gamma }}=S_{\mathbf{\alpha }}S_{\mathbf{\beta }},$ for a fixed $%
\underline{\mathbf{y}}$ $(\underline{\mathbf{y}}=\underline{\mathbf{a}},$ or 
$\underline{\mathbf{y}}=\underline{\mathbf{b}},$ or $\underline{\mathbf{y}}=%
\underline{\mathbf{c}})$ the four matrix elements $H_{\underline{\mathbf{y}}%
\mathbf{\gamma }},$ similarly to the case $N=1,$ can only assume the values $%
(\pm 1),$ which always lead to Eq.~(\ref{CPFTrigAB}) or Eq.~(\ref{CPFZeta}).
On the other hand, for $N>1$ accidentally it may also happen that the CPF
correlation vanishes. This occur because we assumed that the incoherent
degrees of freedom are stationary, which implies $\sum_{\mathbf{\alpha
,\beta }}[(1|\mathbb{F}_{\mathbf{\alpha }}(\tau )\mathbb{F}_{\mathbf{\beta }%
}(t)|q_{0})-(1|\mathbb{F}_{\mathbf{\alpha }}(\tau )|q_{t})(1|\mathbb{F}_{%
\mathbf{\beta }}(t)|q_{0})]=0.$ Thus, when $H_{\underline{\mathbf{y}}\mathbf{%
\gamma }}=1,$ it follows that $C_{pf}(t,\tau )|_{y}=0$ [see Eq.~(\ref{CPFSol}%
)]. These accidental cases can always be surpassed by considering arbitrary
measurement operators written as linear combinations of the Pauli strings. 
%figura1%figura%figura%figura%figurav%figura%figura%figura%figura%figura%figura%figura%figura%figura%figurav%figura%figura%figura%figura%figura
%figura%figura%figura%figura%figurav%figura%figura%figura%figura%figura%figura%figura%figura%figura%figurav%figura%figura%figura%figura%figura
\begin{figure}[tbp]
\includegraphics[bb=44 870 720
1155,angle=0,width=8.5cm]{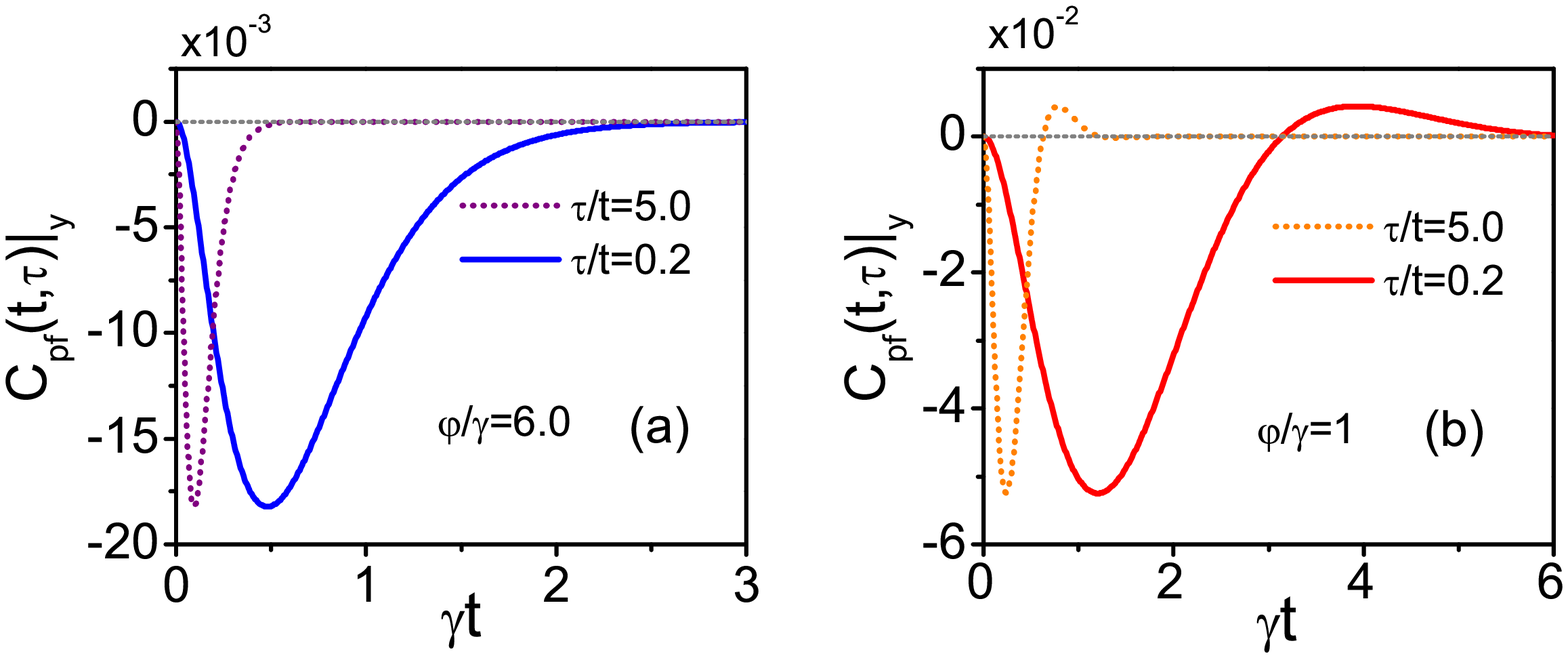} 
\caption{CPF correlation [Eq.~(\protect\ref{CPFTrigAB})] corresponding to
the eternal non-Markovian trigonometric model [Eq.~(\protect\ref%
{RhoTrigonometric})], for different values of $\protect\varphi /\protect%
\gamma $ and measurement time-interval relations $\protect\tau /t.$ In all
cases, the system initial condition is such that $\langle x\rangle =0.$}
\end{figure}
%figura%figura%figura%figura%figurav%figura%figura%figura%figura%figura%figura%figura%figura%figura%figurav%figura%figura%figura%figura%figura
%figura%figura%figura%figura%figurav%figura%figura%figura%figura%figura%figura%figura%figura%figura%figurav%figura%figura%figura%figura%figura

As an example, we consider a bipartite case where $S_{\underline{\mathbf{a}}%
}=\sigma _{1}^{x}\sigma _{2}^{x},$ $S_{\underline{\mathbf{b}}}=\sigma
_{1}^{y}\sigma _{2}^{y},$ and $S_{\underline{\mathbf{c}}}=\sigma
_{1}^{z}\sigma _{2}^{z}.$ The CPF correlation Eq.~(\ref{CPFTrigAB}) is
obtained when the three measurement are defined by any of the bipartite
operators $S_{\underline{\mathbf{m}}}=(\sigma _{1}^{x}\sigma _{2}^{z}), $ $%
(\sigma _{1}^{y}\sigma _{2}^{z}),$ $(\sigma _{1}^{z}\sigma _{2}^{x}),$ $%
(\sigma _{1}^{z}\sigma _{2}^{y}),$ Eq.~(\ref{CPFZeta}) is obtained when $S_{%
\underline{\mathbf{m}}}=(\sigma _{1}^{x}\sigma _{2}^{y}),(\sigma
_{1}^{y}\sigma _{2}^{x}),$ while $C_{pf}(t,\tau )|_{y}=0$ when $S_{%
\underline{\mathbf{m}}}=(\sigma _{1}^{x}\sigma _{2}^{x}),$ $(\sigma
_{1}^{y}\sigma _{2}^{y}),$ $(\sigma _{1}^{z}\sigma _{2}^{z}).$

\section{Summary and conclusions}

We studied a class of solvable multipartite non-Markovian master equations
where the system consists of an arbitrary number of qubits and whose
structure is written in terms of arbitrary multipartite Pauli coupling
terms. Starting from a local-in-time representation of the evolution, we
found the explicit solution for the system density matrix, which in turn
allowed us to formulate the constraints that time-dependent rates must obey
in order to guarantee the completely positive condition of the solution map.

We also found explicit analytical expressions for the time-dependent rates
associated to a given evolution. Their sign (positive or negative) can be
used as an indicator of non-Markovianity. Memory effects were also
characterized by operational methods, where a CPF correlation defined by a
set of three consecutive system measurements becomes a memory witness. We
showed that this quantity can be obtained in an exact way for arbitrary
measurement processes and arbitrary interaction with incoherent degrees of
freedom.

As application of the previous results, we presented simple underlying
dynamics that lead to the phenomenon of eternal non-Markovianity, that is,
multipartite dynamics where some rates depart at all times from that of a
Markovian regime. Both hyperbolic and trigonometric cases were established,
characterized by a rate that is negative at all times or that develops
periodical divergences. Even when these features develop, the CPF
correlation is always a smooth function.

In the Appendices we found the rates associated to different underlying
memory mechanisms such as a mapping with a classical master equation,
stochastic Hamiltonians and statistical superpositions of Markovian
dynamics. We showed that under particular conditions different mechanisms
may lead to the same time-dependent rates. Nevertheless, these accidental
degeneracies do not occur in general. We also found that the phenomenon of
eternal non-Markovianity becomes quite common in multipartite dynamics.

The class of models we studied here provides a useful solvable framework for
studying quantum non-Markovianity in multipartite settings. This allows to
formulate a wide range of well-behaved multipartite non-Markovian master
equations. The study of diverse memory witness can be tackled starting from
here. Our results also lead to interesting questions such as determining
which kind of underlying dynamics can be associated to an arbitrary
non-Markovian multipartite Pauli evolution.

\section*{Acknowledgments}

AAB acknowledges support from CONICET, Argentina. JPG acknowledges financial
support from EPSRC Grant no.\ EP/R04421X/1 and the Leverhulme Trust Grant
No. RPG-2018-181.

\appendix

\section{Time-convoluted approach}

Instead of the local-in-time formulation defined by Eq.~(\ref{class}),
alternatively one may start with a time convoluted evolution%
\begin{equation}
\frac{d}{dt}\rho _{t}=\mathcal{L}[\rho _{t}]=\sum_{\substack{ \mathbf{a}  \\ 
\mathbf{a\neq 0}}}\int_{0}^{t}dt^{\prime }k_{\mathbf{a}}(t-t^{\prime })(S_{%
\mathbf{a}}\rho _{t^{\prime }}S_{\mathbf{a}}-\rho _{t^{\prime }}),
\end{equation}%
where the set of time-dependent kernels $\{k_{\mathbf{a}}(t)\}$ must to be
constrained such that the solution map is CP. Similarly to Sec. II, by
defining the kernel $k_{\mathbf{0}}(t)\equiv -\sum_{\mathbf{a\ (a\neq 0)}}k_{%
\mathbf{a}}(t),$ here the weights of the solution (\ref{Solution}) can be
written as%
\begin{equation}
p_{t}^{\mathbf{a}}=\frac{1}{4^{N}}\sum_{\mathbf{b}}H_{\mathbf{ab}}\lambda _{%
\mathbf{b}}(t),
\end{equation}%
where the coefficients $\lambda _{\mathbf{b}}(t)$ obey the evolution%
\begin{equation}
\frac{d}{dt}\lambda _{\mathbf{b}}(t)=\int_{0}^{t}dt^{\prime }k_{\mathbf{b}%
}(t-t^{\prime })\lambda _{\mathbf{b}}(t^{\prime }).
\end{equation}%
The inverse relations for determining the kernels $\{k_{\mathbf{a}}(t)\}$ a
function of probabilities $\{p_{\mathbf{a}}(t)\}$ can be written in a
Laplace domain $[f(z)=\int_{0}^{\infty }dte^{-zt}f(t)]$ as%
\begin{equation}
k_{\mathbf{a}}(z)=\frac{z\lambda _{\mathbf{a}}(z)-1}{\lambda _{\mathbf{a}}(z)%
},\ \ \ \ \ \ \ \lambda _{\mathbf{a}}(z)=\sum_{\mathbf{b}}H_{\mathbf{ab}}p_{%
\mathbf{b}}(z).
\end{equation}

\section{Non-Markovian underlying mechanisms}

Here, we consider different mechanisms that lead to memory effects. The
present analysis provides nontrivial multipartite extensions of some results
developed in Ref.~\cite{megier} for the case $N=1.$

\subsection{Mapping with a classical Markovian master equation}

The solution map [Eq.~(\ref{Solution})] is defined by a set of normalized
probabilities $\{p_{t}^{\mathbf{a}}\}.$ It is possible to formulate an
underlying mechanism such that $\{p_{t}^{\mathbf{a}}\}$ correspond to the
solution of an arbitrary Markovian classical master equation with $4^{N}$
different states.

We assume that the system density matrix interacts with an incoherent system
whose states, in contrast to Eq.~(\ref{IncoherentStructure}), can be put in
one-to-one correspondence with the Pauli string vectors $\{\mathbf{a}\}.$
Therefore, the system density matrix $\rho _{t}$ can be written in terms of
a set of auxiliary states $\{\rho _{t}^{\mathbf{a}}\}$ \cite{lindbladrate}
such that%
\begin{equation}
\rho _{t}=\sum_{\mathbf{a}}\rho _{t}^{\mathbf{a}}.  \label{Suma}
\end{equation}%
The evolution of the auxiliary states is Markovian and involves coupling
between all of them. We write%
\begin{equation}
\frac{d}{dt}\rho _{t}^{\mathbf{a}}=-\sum_{\substack{ \mathbf{b}  \\ \mathbf{b%
}\neq \mathbf{a}}}\phi _{\mathbf{ba}}\rho _{t}^{\mathbf{a}}+\sum_{\substack{ 
\mathbf{b}  \\ \mathbf{b}\neq \mathbf{a}}}\phi _{\mathbf{ab}}S_{\mathbf{a}%
}S_{\mathbf{b}}\rho _{t}^{\mathbf{b}}S_{\mathbf{b}}S_{\mathbf{a}}.
\label{LindbladRate}
\end{equation}%
Here, $\{\phi _{\mathbf{ba}}\}$ are arbitrary rates. The stochastic
interpretation of this equation is quite simple. Whenever the incoherent
system undergoes the transition $\mathbf{b}\rightarrow \mathbf{a},$ the
quantum system undergoes the transformation $\rho \rightarrow S_{\mathbf{a}%
}S_{\mathbf{b}}\rho S_{\mathbf{b}}S_{\mathbf{a}}.$ Between transition the
system is frozen. The average system dynamics is given by Eq.~(\ref%
{LindbladRate}), where $\rho _{t}^{\mathbf{a}}$\ corresponds to the
conditional system state given that the incoherent one is in the state
associated to $\mathbf{a}.$

It is simple to check that the solutions $\{\rho _{t}^{\mathbf{a}}\}$ of Eq.
(\ref{LindbladRate}) can be written as%
\begin{equation}
\rho _{t}^{\mathbf{a}}=p_{t}^{\mathbf{a}}(S_{\mathbf{a}}\rho _{0}S_{\mathbf{a%
}}),  \label{RhoA}
\end{equation}%
where the weights $p_{t}^{\mathbf{a}}$ must to fulfill the classical master
equation%
\begin{equation}
\frac{d}{dt}p_{t}^{\mathbf{a}}=-\sum_{\substack{ \mathbf{b}  \\ \mathbf{b}%
\neq \mathbf{a}}}\phi _{\mathbf{ba}}p_{t}^{\mathbf{a}}+\sum_{\substack{ 
\mathbf{b}  \\ \mathbf{b}\neq \mathbf{a}}}\phi _{\mathbf{ab}}p_{t}^{\mathbf{b%
}}.  \label{master}
\end{equation}%
Consequently, from Eqs.~(\ref{Suma}) and (\ref{RhoA}) we recover the
solution Eq.~(\ref{Solution}) $[\rho _{t}=\sum_{\mathbf{a}}p_{t}^{\mathbf{a}%
}(S_{\mathbf{a}}\rho _{0}S_{\mathbf{a}})],$ where the probabilities $%
\{p_{t}^{\mathbf{a}}\}$ fulfill a the classical master equation~(\ref{master}%
). For consistence, its initial condition must be $p_{0}^{\mathbf{a}}=\delta
_{\mathbf{a,0}}.$

\textit{Particular case}: Given that Eq.~(\ref{master}) is arbitrary, it is
not possible to find a general expression for the rates $\{\gamma _{t}^{%
\mathbf{a}}\}$ [Eq.~(\ref{SolRates})] in terms of the underlying ones $%
\{\phi _{\mathbf{ba}}\}.$ Nevertheless, this mapping can be performed, for
example, when \ Eq.~(\ref{master}) assumes the form%
\begin{equation}
\frac{d}{dt}p_{t}^{\mathbf{0}}=-\phi p_{t}^{\mathbf{0}}+\varphi \sum 
_{\substack{ \mathbf{a}  \\ \mathbf{a\neq 0}}}p_{t}^{\mathbf{a}},\ \ \ \ \ 
\frac{d}{dt}p_{t}^{\mathbf{a}}=-\varphi p_{t}^{\mathbf{a}}+x_{\mathbf{a}%
}\phi p_{t}^{\mathbf{0}},
\end{equation}%
where $\phi $ and $\varphi $ are arbitrary rates and the weights $\{x_{%
\mathbf{a}}\}$ satisfies $\sum_{\mathbf{a(a}\neq \mathbf{0)}}x_{\mathbf{a}%
}=1.$ The probabilities, with initial condition $p_{0}^{\mathbf{a}}=\delta _{%
\mathbf{a,0}},$ can be written as%
\begin{equation}
p_{t}^{\mathbf{a}}=p_{\infty }^{\mathbf{a}}[1-\exp (-\Phi t)]+\delta _{%
\mathbf{a,0}}\exp (-\Phi t),  \label{ProbSimple}
\end{equation}%
where $\Phi \equiv (\phi +\varphi ),$ and the stationary values are%
\begin{equation}
p_{\infty }^{\mathbf{0}}=\frac{\varphi }{\phi +\varphi },\ \ \ \ \ \
p_{\infty }^{\mathbf{a}}=\frac{x_{\mathbf{a}}\phi }{\phi +\varphi }.
\label{station}
\end{equation}%
From the solutions (\ref{ProbSimple}), the general expression (\ref{SolRates}%
), after some calculations steps \cite{tanh}, lead to%
\begin{equation}
\gamma _{t}^{\mathbf{a}}=\frac{1}{4^{N}}\sum_{\mathbf{b}}\frac{\Phi }{2}H_{%
\mathbf{ab}}\left[ \tanh \left( \frac{t\Phi }{2}+\zeta _{\mathbf{b}}\right)
-1\right] ,  \label{GamaEternalMaster}
\end{equation}%
where the parameters are%
\begin{equation}
\zeta _{\mathbf{b}}\equiv \frac{1}{2}\ln \left( \frac{h_{\infty }^{\mathbf{b}%
}}{1-h_{\infty }^{\mathbf{b}}}\right) ,\ \ \ \ \ \ \ \ \ h_{\infty }^{%
\mathbf{b}}\equiv \sum_{\mathbf{c}}H_{\mathbf{bc}}p_{\infty }^{\mathbf{c}}.
\label{Bufanda}
\end{equation}%
It is simple to check that, due to probability normalization, $h_{\infty }^{%
\mathbf{0}}=1.$ Hence, in Eq.~(\ref{GamaEternalMaster}) the term with $%
\mathbf{b=0}$ cancels out. Furthermore if $h_{\infty }^{\mathbf{b}}=0,$ it
follows $\tanh (t\Phi /2+\zeta _{\mathbf{b}})\rightarrow -1.$ In general,
the time dependence of the rate $\gamma _{t}^{\mathbf{a}}$ arise from a
linear combination of hyperbolic tangent functions with coefficient that are 
$\pm 1.$ Thus, in general some rates can be negative at any time.

\subsection{Stochastic Hamiltonians}

We consider a stochastic evolution, where the system wave vector $|\psi
_{t}\rangle $\ is driven by a stochastic Hamiltonian,%
\begin{equation}
\frac{d|\psi _{t}\rangle }{dt}=-iH_{st}|\psi _{t}\rangle =-i\frac{1}{2}\xi
_{t}^{\mathbf{\alpha }}S_{\mathbf{\alpha }}|\psi _{t}\rangle .
\end{equation}%
The Hamiltonian $H_{st}$ is characterized by a noise with an arbitrary
statistics but null average $\langle \langle \xi _{t}^{\mathbf{\alpha }%
}\rangle \rangle =0.$ The index $\mathbf{\alpha \leftrightarrow \alpha }_{t}$
run overs all possible Pauli strings. Its time variation is very slow such
that over a single realization it can be considered as a frozen parameter.
Thus, the average state $\rho _{t}^{\mathbf{\alpha }}=\langle \langle |\psi
_{t}\rangle \langle \psi _{t}|\rangle \rangle $ for a given $\mathbf{\alpha }
$ reads $\rho _{t}^{\mathbf{\alpha }}=(1/2)[1+G_{t}^{\mathbf{\alpha }}]\rho
_{0}+(1/2)[1-G_{t}^{\mathbf{\alpha }}](S_{\mathbf{\alpha }}\rho _{0}S_{%
\mathbf{\alpha }}),$ where%
\begin{equation}
G_{t}^{\mathbf{\alpha }}\equiv \Big{\langle}\Big{\langle}\exp \Big{(}%
i\int_{0}^{t}dt^{\prime }\xi _{t^{\prime }}^{\mathbf{\alpha }}\Big{)}%
\Big{\rangle}\Big{\rangle},  \label{Caracteristica}
\end{equation}%
is the characteristic noise function for a given $\mathbf{\alpha .}$ After
averaging this parameter, the system state can be written as $\rho
_{t}=\sum_{\mathbf{\alpha },(\mathbf{\alpha }\neq \mathbf{0})}x_{\mathbf{%
\alpha }}\rho _{t}^{\mathbf{\alpha }},$ where $\sum_{\mathbf{\alpha },(%
\mathbf{\alpha }\neq \mathbf{0})}x_{\mathbf{\alpha }}=1.$ The parameters $%
\{x_{\mathbf{\alpha }}\}$ correspond to the statistical weight of each Pauli
string during the variation of the coefficient $\mathbf{\alpha .}$ It is
straightforward to check that $\rho _{t}=\sum_{\mathbf{a}}p_{t}^{\mathbf{a}%
}(S_{\mathbf{a}}\rho _{0}S_{\mathbf{a}}),$ which recovers Eq.~(\ref{Solution}%
) with%
\begin{equation}
p_{t}^{\mathbf{0}}=\frac{1}{2}(1+\sum_{\substack{ \mathbf{a}  \\ \mathbf{a}%
\neq \mathbf{0}}}x_{\mathbf{a}}G_{t}^{\mathbf{a}}),\ \ \ \ \ \ p_{t}^{%
\mathbf{a}}=\frac{x_{\mathbf{a}}}{2}(1-G_{t}^{\mathbf{a}}).
\label{ProbStochasticHamiltonian}
\end{equation}%
Similarly to the previous model, it is not possible to find a general simple
expression for the rates $\gamma _{t}^{\mathbf{a}}$ in terms of these
probabilities. Manageable expressions arise in the following situations.

\textit{Particular cases}: If the noise is the same for all
\textquotedblleft directions\textquotedblright\ $G_{t}^{\mathbf{a}}=G_{t},$
from Eqs.~(\ref{SolRates}) and (\ref{ProbStochasticHamiltonian}), after some
algebra \cite{tanh}, we get the rates%
\begin{equation}
\gamma _{t}^{\mathbf{a}}=\frac{1}{4^{N}}\sum_{\mathbf{b}}\frac{\dot{g}_{t}}{2%
}H_{\mathbf{ab}}\left[ \tanh \left( \frac{g_{t}}{2}+\zeta _{\mathbf{b}%
}\right) -1\right] ,  \label{RateHst}
\end{equation}%
where the scalar functions read%
\begin{equation}
g_{t}=\ln (1/G_{t}),\ \ \ \ \ \ \ \dot{g}_{t}=-\frac{1}{G_{t}}\frac{dG_{t}}{%
dt},  \label{gminuscula}
\end{equation}%
and where $\zeta _{\mathbf{b}}$ is defined by Eq.~(\ref{Bufanda}) with,
instead of Eq.~(\ref{station}), with $p_{\infty }^{\mathbf{0}}=1/2,$ and $%
p_{\infty }^{\mathbf{a}}=x_{\mathbf{a}}/2.$

For a stationary \textit{Gaussian white noise}, where $\langle \langle \xi
_{t}\xi _{t^{\prime }}\rangle \rangle =\Phi \delta (t-t^{\prime }),$ Eq.~(%
\ref{Caracteristica}) becomes $G(t)=\exp (-\Phi t).$ It is simple to check
that in this situation Eq.~(\ref{RateHst}) recovers the solution\ (\ref%
{GamaEternalMaster}) of the previous model with $\varphi =\phi .$ This
results show that there are different underlying models that may lead to the
same system density matrix evolution. This degeneracy is not universal and
clearly depends on the underlying parameters.

For a stationary \textit{symmetric dichotomic noise} with amplitude $A$ and
switching rate $\eta ,$ the characteristic noise function [Eq.~(\ref%
{Caracteristica})] is%
\begin{equation}
G_{t}=e^{-\eta t}[\cosh (\chi t)+\frac{\eta }{\chi }\sinh (\chi t)],\ \ \ \
\ \chi \equiv \sqrt{\eta ^{2}-A^{2}}.
\end{equation}%
In contrast to the previous cases, here the rates defined by Eq.~(\ref%
{RateHst}) may develop divergences. In fact, the functions (\ref{gminuscula}%
) become%
\begin{equation}
g_{t}=\ln (1/G_{t}),\ \ \ \ \ \ \ \dot{g}_{t}=\frac{A^{2}}{\eta +\chi
\lbrack 1/\tanh (\chi t)]}.
\end{equation}%
Hence, divergent rates are found whenever $\eta <A.$

\subsection{Statistical mixtures of Markovian evolutions}

Departures with respect to a Markovian regime emerge whenever the system
evolution is written as the statistical superposition of different Markovian
propagators. Hence, we write%
\begin{equation}
p_{t}^{\mathbf{a}}=\sum_{k=1}^{n}q_{k}p_{k}^{\mathbf{a}}(t),
\label{StatisticalSuperposition}
\end{equation}%
where $\{q_{k}\}$ are normalized positive weights $(\sum_{k=1}^{n}q_{k}=1),$
and each set of probabilities $\{p_{k}^{\mathbf{a}}(t)\}$ is associated to a
Markovian solution of Eq.~(\ref{class}) with time-independent positive rates 
$\{\gamma _{k}^{\mathbf{a}}\}.$

From Eq.~(\ref{SolRates}), the non-Markovian evolution is characterized by
the rates%
\begin{equation}
\gamma _{t}^{\mathbf{a}}=\frac{1}{4^{N}}\sum_{\mathbf{b}}H_{\mathbf{ab}}%
\frac{\sum_{k=1}^{n}q_{k}\mu _{k}^{\mathbf{b}}\exp (t\mu _{k}^{\mathbf{b}})}{%
\sum_{k^{\prime }=1}^{n}q_{k^{\prime }}\exp (t\mu _{k^{\prime }}^{\mathbf{b}%
})}.  \label{GamaRandomSuper}
\end{equation}%
where $\mu _{k}^{\mathbf{b}}$ are eigenvalues of the $k$-Markovian dynamics, 
$\mu _{k}^{\mathbf{b}}=\sum_{\mathbf{c}}H_{\mathbf{bc}}\gamma _{k}^{\mathbf{c%
}}.$ The specific properties of these rates strongly depend on the
considered Markovian evolutions and statistic weights.

\textit{Particular cases}: In the \textit{two-state case,} $n=2,$ the
probabilities are $p_{t}^{\mathbf{a}}=q_{1}p_{1}^{\mathbf{a}}(t)+q_{2}p_{2}^{%
\mathbf{a}}(t),$ where each solution is associated to the rates $\gamma
_{1}^{\mathbf{a}}$ and $\gamma _{2}^{\mathbf{a}},$ and $q_{1}+q_{2}=1.$ From
Eq.~(\ref{GamaRandomSuper}), after some algebra \cite{tanh}, we get 
\begin{equation}
\gamma _{t}^{\mathbf{a}}=\frac{1}{2}(\gamma _{1}^{\mathbf{a}}+\gamma _{2}^{%
\mathbf{a}})+\frac{1}{4^{N}}\sum_{\mathbf{b}}H_{\mathbf{ab}}\Delta _{\mathbf{%
b}}\tanh (t\Delta _{\mathbf{b}}+\zeta ),  \label{GamaTwoMixture}
\end{equation}%
where the parameters are%
\begin{equation}
\Delta _{\mathbf{b}}\equiv \frac{1}{2}\sum_{\mathbf{c}}H_{\mathbf{bc}%
}(\gamma _{1}^{\mathbf{c}}-\gamma _{2}^{\mathbf{c}}),\ \ \ \ \ \ \ \zeta
\equiv \frac{1}{2}\ln (\frac{q_{1}}{q_{2}}).
\end{equation}%
In this case, many rates may also be negative at all times (see next
section).

In the other extreme, a \textit{continuos-state case} can be considered.
Thus, Eq.~(\ref{StatisticalSuperposition}) is rewritten as%
\begin{equation}
p_{t}^{\mathbf{a}}=\frac{1}{4^{N}}\sum_{\mathbf{b}}H_{\mathbf{ab}}%
\Big{\langle}\prod_{\mathbf{c}}\exp (tH_{\mathbf{bc}}\gamma ^{\mathbf{c}})%
\Big{\rangle},
\end{equation}%
where we used the explicit expression (\ref{ProbSol}) and the replacement $%
\sum_{k=1}^{n}q_{k}\rightarrow \left\langle \cdots \right\rangle .$ The
symbol $\left\langle \cdots \right\rangle $ denotes an average over the set
of random rates $\{\gamma ^{\mathbf{c}}\},$ each \textquotedblleft
realization\textquotedblright\ defining a Markov solution. Assuming that all
rates are independent random variables it follows that $\left\langle \cdots
\right\rangle \rightarrow \int_{0}^{\infty }d\gamma ^{\mathbf{c}}\cdots
P(\gamma ^{\mathbf{c}}),$ where $P(\gamma ^{\mathbf{c}})$ is the
corresponding probability density. By assuming an \textit{exponential
probability density} $P(\gamma ^{\mathbf{c}})=\tau _{\mathbf{c}}\exp
(-\gamma ^{\mathbf{c}}\tau _{\mathbf{c}}),$ by using that $\gamma ^{\mathbf{0%
}}=-\sum_{\mathbf{c}(\mathbf{c}\neq \mathbf{0})}\gamma ^{\mathbf{c}},$ [see
Eq.~(\ref{GamaCero})] we get 
\begin{equation}
p_{t}^{\mathbf{a}}=\frac{1}{4^{N}}\sum_{\mathbf{b}}H_{\mathbf{ab}}\prod 
_{\substack{ \mathbf{c}  \\ \mathbf{c}\neq \mathbf{0}}}\frac{\tau _{\mathbf{c%
}}}{\tau _{\mathbf{c}}+(1-H_{\mathbf{bc}})t},
\end{equation}%
where we have used that $H_{\mathbf{b0}}=1.$ From Eq.~(\ref{SolRates}), the
corresponding rates associated to the non-Markovian evolution are%
\begin{equation}
\gamma _{t}^{\mathbf{a}}=-\frac{1}{4^{N}}\sum_{\mathbf{b}}H_{\mathbf{ab}%
}\sum _{\substack{ \mathbf{c}  \\ \mathbf{c}\neq \mathbf{0}}}\frac{(1-H_{%
\mathbf{bc}})}{\tau _{\mathbf{c}}+(1-H_{\mathbf{bc}})t}.
\end{equation}

We notice that both $\{p_{t}^{\mathbf{a}}\}$ and $\{\gamma _{t}^{\mathbf{a}%
}\}$ develop a power-law behavior. In spite of this feature the rates are
positive at all times, $\gamma _{t}^{\mathbf{a}}>0$ $(\mathbf{a}\neq \mathbf{%
0}).$ While most of the memory witnesses \cite{BreuerReview,plenioReview}\
associate this property to a Markovian regime, from operational approaches
it is possible to detect and infer the presence of memory effects~\cite%
{budiniCPF,BIF}.

\section{Bipartite and tripartite eternal non-Markovian evolutions}

Besides the previous examples, the developed approach allow us to show that
master equations characterized by eternal non-Markovian effects are quite
common for multipartite systems. As an example, we consider the statistical
superposition of two different Markovian dynamics characterized by the
rates\ $\gamma _{1}^{\mathbf{a}}$ and $\gamma _{2}^{\mathbf{a}}$ and equal
weights $[q_{1}=q_{2}$ in Eq.~(\ref{GamaTwoMixture})]. Taking $\gamma _{1}^{%
\mathbf{a}}=\gamma (\delta _{\mathbf{a},\underline{\mathbf{a}}}-\delta _{%
\mathbf{a},\mathbf{0}}),$ and $\gamma _{2}^{\mathbf{a}}=\gamma (\delta _{%
\mathbf{a},\underline{\mathbf{b}}}-\delta _{\mathbf{a},\mathbf{0}}),$ and
using that $(H_{\mathbf{\alpha }\underline{\mathbf{a}}}-H_{\mathbf{\alpha }%
\underline{\mathbf{b}}})/2=(\pm 1,0),$ and $H_{\mathbf{\alpha }\underline{%
\mathbf{a}}}H_{\mathbf{\alpha }\underline{\mathbf{b}}}=H_{\mathbf{\alpha }%
\underline{\mathbf{c}}},$ from Eq.~(\ref{GamaTwoMixture}) we recover the
rates defined in Eq. (\ref{Tanh}). When each (vectorial) rate involves
different Pauli channels more complex expressions are obtained.

As a first example, take a \textit{bipartite} system $(N=2)$ with 
\begin{subequations}
\begin{eqnarray}
\gamma _{1}^{\mathbf{a}} &=&\gamma (\delta _{\mathbf{a},10}+\delta _{\mathbf{%
a},01}-2\delta _{\mathbf{a},00}), \\
\gamma _{2}^{\mathbf{a}} &=&\gamma (\delta _{\mathbf{a},20}+\delta _{\mathbf{%
a},02}-2\delta _{\mathbf{a},00}).
\end{eqnarray}%
Thus, each dynamics is defined by a local (single) dephasing local mechanism
acting alternatively in $x$- and $y$-directions. From Eq.~(\ref%
{GamaTwoMixture}) we obtain 
\end{subequations}
\begin{equation}
\gamma _{t}^{\mathbf{a}_{0}}=\frac{1}{2}\gamma ,\ \ \ \ \ \ \ \gamma _{t}^{%
\mathbf{a}_{\pm }}=\pm \frac{1}{4}\gamma \tanh (2\gamma t),  \label{GamaBip}
\end{equation}%
where $\mathbf{a}_{0}$ and $\mathbf{a}_{\pm }$ correspond to the following
Pauli strings, $\mathbf{a}_{0}=(10),\ (01),$\ $(20),$ $(02),$ and $\mathbf{a}%
_{+}=(11),\ (22),$ while $\mathbf{a}_{-}=(30),$\ $(03),$\ $(12),$\ $(21).$
Furthermore,%
\begin{equation*}
\gamma _{t}^{33}=-\frac{\gamma }{4}[2\tanh (\gamma t)-\tanh (2\gamma
t)]=-2\gamma \frac{\sinh ^{4}(\gamma t)}{\sinh (4\gamma t)},
\end{equation*}%
while $\gamma _{t}^{\mathbf{a}}=0$ if $\mathbf{a}\neq (\mathbf{a}_{0},%
\mathbf{a}_{+},\mathbf{a}_{-}).$ There are eleven non-null rates out of the
fifteen possible ones, five of them being negative at all times.

As a second example we consider a \textit{tripartite} system $(N=3),$ where 
\begin{subequations}
\begin{eqnarray}
\gamma _{1}^{\mathbf{a}} &=&\gamma (\delta _{\mathbf{a},110}+\delta _{%
\mathbf{a},101}+\delta _{\mathbf{a},011}-3\delta _{\mathbf{a},000}), \\
\gamma _{2}^{\mathbf{a}} &=&\gamma (\delta _{\mathbf{a},220}+\delta _{%
\mathbf{a},202}+\delta _{\mathbf{a},022}-3\delta _{\mathbf{a},000}).
\end{eqnarray}%
Hence, each Markovian evolution correspond to dephasing in $x$- and $y$%
-directions but now considering all pairs of bipartite dephasing operators.
From Eq.~(\ref{GamaTwoMixture}) we get 
\end{subequations}
\begin{subequations}
\label{GamaTrip}
\begin{eqnarray}
\gamma _{t}^{\mathbf{a}_{+}} &=&\frac{1}{4}\gamma \lbrack 2+\tanh (2\gamma
t)], \\
\gamma _{t}^{\mathbf{a}_{-}} &=&-\frac{1}{4}\gamma \tanh (2\gamma t),
\end{eqnarray}%
where $\mathbf{a}_{\pm }$ correspond to the following Pauli strings, $%
\mathbf{a}_{+}=(110),\ (101),$\ $(011),\ (220),\ (202),\ (022),$ while $%
\mathbf{a}_{-}=(330),$\ $(303),$\ $(033),$\ $(123),$\ $(132),$\ $(213),$\ $%
(231),$\ $(312),$\ $(321),$ and $\gamma _{t}^{\mathbf{a}}=0$ if $\mathbf{a}%
\neq \mathbf{a}_{+},\mathbf{a}_{-}.$ In this case, out of sixty-three
possible rates, fifteen are non-null, nine of them being negative at all
times.

\section{CPF correlation calculus}

For a system coupled to incoherent degrees of freedom [Eq.~(\ref%
{IncoherentStructure})], the (bipartite) system-environment state $\rho
_{t}^{se}=\sum_{\mathbf{h}}\rho _{t}^{\mathbf{h}}|\mathbf{h}),$ from Eqs.~(%
\ref{RhoAux}) and (\ref{gVectorial}), reads 
\end{subequations}
\begin{equation}
\rho _{t}^{se}=\sum_{\mathbf{\alpha }}(S_{\mathbf{\alpha }}\rho _{0}S_{%
\mathbf{\alpha }})\ \mathbb{F}_{\mathbf{\alpha }}(t)|q_{0}).
\end{equation}%
This evolution defines the system-environment dynamics between measurements.
The measurement of operator $S_{\underline{\mathbf{m}}}$ [Eq.~(\ref%
{Observables})] leads to the transformation $\rho ^{se}=\sum_{\mathbf{h}%
}\rho ^{\mathbf{h}}|\mathbf{h})\rightarrow |m\rangle \langle m||q_{m}),$
where $|q_{m})=\sum_{\mathbf{h}}\ \langle m|\rho _{t}^{\mathbf{h}}|m\rangle /%
\mathrm{Tr}[\langle m|\rho _{t}^{\mathbf{h}}|m\rangle ]|\mathbf{h}).$ With
these ingredients, the calculation of the joint probability can be performed
in a standard way. We get,%
\begin{equation}
\frac{P(z,y,x)}{P(x)}=\sum_{\mathbf{\alpha ,\beta }}|\langle z|\sigma _{%
\mathbf{\alpha }}|y\rangle |^{2}|\langle y|S_{\mathbf{\beta }}|x\rangle
|^{2}(1|\mathbb{F}_{\mathbf{\alpha }}(\tau )\mathbb{F}_{\mathbf{\beta }%
}(t)|q_{0}),  \label{P3Joint}
\end{equation}%
where $P(x)=\langle x|\rho _{0}|x\rangle $ and $(1|\equiv \sum_{\mathbf{h}}(%
\mathbf{h}|.$ This result is valid for arbitrary Hermitian system
observables.

Using Bayes rule, the conditional probabilities that define the CPF
correlation [Eq.~(\ref{CPFexplicit})] can be written as $%
P(z,x|y)=P(z,y,x)/P(y),$ where $P(y)=\sum_{z,x}P(z,y,x).$ Furthermore, $%
P(z|y)=\sum_{x}P(z,x|y),$ and $P(x|y)=\sum_{z}P(z,x|y).$ From Eq. (\ref%
{P3Joint}), and using 
\begin{subequations}
\begin{eqnarray}
\sum_{z}z|\langle z|S_{\mathbf{\alpha }}|y\rangle |^{2} &=&\langle y|S_{%
\mathbf{\alpha }}S_{\underline{\mathbf{z}}}S_{\mathbf{\alpha }}|y\rangle , \\
\sum_{x}x|\langle y|S_{\mathbf{\beta }}|x\rangle |^{2}P(x) &=&\langle y|S_{%
\mathbf{\beta }}S_{\underline{\mathbf{x}}}\rho _{\underline{\mathbf{x}}}S_{%
\mathbf{\beta }}|y\rangle , \\
\sum_{x}|\langle y|S_{\mathbf{\beta }}|x\rangle |^{2}P(x) &=&\langle y|S_{%
\mathbf{\beta }}\rho _{\underline{\mathbf{x}}}S_{\mathbf{\beta }}|y\rangle ,
\end{eqnarray}%
where the system state $\rho _{\underline{\mathbf{x}}}$ is 
\end{subequations}
\begin{equation}
\rho _{\underline{\mathbf{x}}}\equiv \sum_{x}P(x)\ |x\rangle \langle
x|=\sum_{x}\langle x|\rho _{0}|x\rangle \ |x\rangle \langle x|,
\end{equation}%
the CPF correlation can be written as%
\begin{equation}
C_{pf}(t,\tau )|_{y}=\frac{1}{P(y)^{2}}\sum_{\mathbf{\alpha ,\beta ,\gamma }%
}\Theta ^{\mathbf{\alpha \beta \gamma }}|_{y}\Lambda _{\mathbf{\alpha \beta
\gamma }}(t,\tau ).  \label{CPFGen}
\end{equation}%
The coefficients $\Theta ^{\mathbf{\alpha \beta \gamma }}|_{y}$ are%
\begin{equation*}
\Theta ^{\mathbf{\alpha \beta \gamma }}|_{y}=\langle y|S_{\mathbf{\alpha }%
}S_{\underline{\mathbf{z}}}S_{\mathbf{\alpha }}|y\rangle \langle y|S_{%
\mathbf{\beta }}S_{\underline{\mathbf{x}}}\rho _{\underline{\mathbf{x}}}S_{%
\mathbf{\beta }}|y\rangle \langle y|\sigma _{\mathbf{\gamma }}\rho _{%
\underline{\mathbf{x}}}S_{\mathbf{\gamma }}|y\rangle ,
\end{equation*}%
while the time-dependence follows from%
\begin{eqnarray*}
\Lambda _{\mathbf{\alpha \beta \gamma }}(t,\tau ) &=&+(1|\mathbb{F}_{\mathbf{%
\alpha }}(\tau )\mathbb{F}_{\mathbf{\beta }}(t)|q_{0})(1|\mathbb{F}_{\mathbf{%
\gamma }}(t)|q_{0}) \\
&&-(1|\mathbb{F}_{\mathbf{\alpha }}(\tau )\mathbb{F}_{\mathbf{\gamma }%
}(t)|q_{0})(1|\mathbb{F}_{\mathbf{\beta }}(t)|q_{0}),
\end{eqnarray*}%
where $|q_{t})=\sum_{\mathbf{\alpha }}\mathbb{F}_{\mathbf{\alpha }%
}(t)|q_{0}),$ and the probability $P(y)$ is%
\begin{equation}
P(y)=\sum_{\mathbf{\alpha }}(1|\mathbb{F}_{\mathbf{\alpha }%
}(t)|q_{0})\langle y|S_{\mathbf{\alpha }}\rho _{\underline{\mathbf{x}}}S_{%
\mathbf{\alpha }}|y\rangle .
\end{equation}

The expression\ (\ref{CPFGen}) is valid for arbitrary observables $\sigma _{%
\underline{\mathbf{m}}}$ [Eq.~(\ref{Observables})]. In general, they can be
written as linear combinations of Pauli strings $S_{\mathbf{a}}.$ Assuming,
for simplicity, that each $S_{\underline{\mathbf{m}}}$ correspond to a
unique Pauli string operator, from Eq.~(\ref{HadaKraus}) it follows the
relations 
\begin{subequations}
\begin{eqnarray}
\langle y|S_{\mathbf{\alpha }}S_{\underline{\mathbf{z}}}\sigma _{\mathbf{%
\alpha }}|y\rangle &=&H_{\mathbf{\alpha }\underline{\mathbf{y}}}\delta _{%
\underline{\mathbf{z}},\mathbf{y}}a_{y}, \\
\langle y|S_{\mathbf{\beta }}S_{\underline{\mathbf{x}}}\rho _{\underline{%
\mathbf{x}}}S_{\mathbf{\beta }}|y\rangle &=&\frac{1}{2^{N}}(H_{\mathbf{\beta 
}\underline{\mathbf{y}}}\delta _{\underline{\mathbf{y}},\underline{\mathbf{x}%
}}a_{y}+\langle x\rangle ), \\
\langle y|S_{\mathbf{\gamma }}\rho _{\underline{\mathbf{x}}}\sigma _{\mathbf{%
\gamma }}|y\rangle &=&\frac{1}{2^{N}}(1+H_{\mathbf{\gamma }\underline{%
\mathbf{y}}}\delta _{\underline{\mathbf{y}},\underline{\mathbf{x}}%
}a_{y}\langle x\rangle ),\ \ \ 
\end{eqnarray}%
where $\langle x\rangle \equiv \mathrm{Tr}[S_{\underline{\mathbf{x}}}\rho _{%
\underline{\mathbf{x}}}].$ By introducing these equalities in Eq.~(\ref%
{CPFGen}), after some algebra we get Eq. (\ref{CPFSol}). Generalization to
arbitrary observables can be worked out in a similar way from Eq.~(\ref%
{CPFGen}).

\end{subequations}


\begin{thebibliography}{99}
\bibitem{alicki} R. Alicki and K. Lendi, \textit{Quantum Dynamical
Semigroups and Applications}, Lecture Notes in Physics \textbf{286}
(Springer, Berlin, 1987).

\bibitem{breuerbook} H. P. Breuer and F. Petruccione, \textit{The theory of
open quantum systems}, (Oxford University Press, 2002).

\bibitem{vega} I. de Vega and D. Alonso, Dynamics of non-Markovian open
quantum systems, Rev. Mod. Phys. \textbf{89}, 015001 (2017).

\bibitem{wiseman} L. Li, M. J. W. Hall, and H. M. Wiseman, Concepts of
quantum non-Markovianity: A hierarchy, Phys. Rep. \textbf{759}, 1 (2018).

\bibitem{BreuerReview} H. P. Breuer, E. M. Laine, J. Piilo, and V. Vacchini,
Colloquium: Non-Markovian dynamics in open quantum systems, Rev. Mod. Phys. 
\textbf{88}, 021002 (2016).

\bibitem{plenioReview} A. Rivas, S. F. Huelga, and M. B. Plenio, Quantum
non-Markovianity: characterization, quantification and detection, Rep. Prog.
Phys. \textbf{77}, 094001 (2014).

\bibitem{wilkie} J. Wilkie, Positivity preserving non-Markovian master
equations, Phys. Rev. E \textbf{62}, 8808 (2000).

\bibitem{barnett} S. M. Barnett and S. Stenholm, Hazards of reservoir
memory, Phys. Rev. A \textbf{64}, 033808 (2001).

\bibitem{budini} A. A. Budini, Stochastic representation of a class of
non-Markovian completely positive evolutions, Phys. Rev. A \textbf{69},
042107 (2004).

\bibitem{cresserJD} S. Daffer, K. Wodkiewicz, J.D. Cresser, and J.K. McIver,
Depolarizing channel as a completely positive map with memory, Phys. Rev. A 
\textbf{70}, 010304(R) (2004).

\bibitem{LocalNonLocal} D. Chru\'{s}ci\'{n}ski and A. Kossakowski,
Non-Markovian Quantum Dynamics: Local versus Nonlocal, Phys. Rev. Lett. 
\textbf{104}, 070406 (2010).

\bibitem{GaussianNoise} A. A. Budini, Quantum systems subject to the action
of classical stochastic fields, Phys. Rev. A \textbf{64}, 052110 (2001); J.
I. Costa-Filho, R. B. B. Lima, R. R. Paiva, P. M. Soares, W. A. M. Morgado,
R. Lo Franco, and D. O. Soares-Pinto, Enabling quantum non-Markovian
dynamics by injection of classical colored noise, Phys. Rev. A \textbf{95},
052126 (2017); Cialdi, C. Benedetti , D. Tamascelli, S. Olivares, M. G. A.
Paris, and B. Vacchini, Experimental investigation of the effect of
classical noise on quantum non-Markovian dynamics, Phys. Rev. A \textbf{100}%
, 052104 (2019); A. Kiely, Exact classical noise master equations:
Applications and connections, Euro Phys. Lett. \textbf{13}4, 10001 (2021).

\bibitem{shabani} A. Shabani and D. A. Lidar, Completely positive
post-Markovian master equation via a measurement approach, Phys. Rev. A 
\textbf{71}, 020101(R) (2005); C. Sutherland, T. A. Brun, and D. A. Lidar,
Non-Markovianity of the post-Markovian master equation, Phys. Rev. A \textbf{%
98}, 042119 (2018).

\bibitem{petruccioneLidarEq} S. Maniscalco and F. Petruccione, Non-Markovian
dynamics of a qubit, Phys. Rev. A \textbf{73}, 012111 (2006).

\bibitem{salo} J. Salo, S. M. Barnett, and S. Stenholm, Non-Markovian
thermalization of a two-level system, Op. Comm. \textbf{259}, 772 (2006).

\bibitem{kossaDariusz} D. Chru\'{s}ci\'{n}ski and A. Kossakowski, Sufficient
conditions for a memory-kernel master equation, Phys. Rev. A \textbf{94},
020103(R) (2016).

\bibitem{lindbladrate} A. A. Budini, Lindblad rate equations, Phys. Rev. A 
\textbf{74}, 053815 (2006).

\bibitem{PostMarkovian} A. A. Budini, Post-Markovian quantum master
equations from classical environment fluctuations, Phys. Rev. E \textbf{89},
012147 (2014).

\bibitem{boltzman} B. Vacchini, Non-Markovian dynamics for bipartite
systems, Phys. Rev. A \textbf{78}, 022112 (2008).

\bibitem{megier} N. Megier, D. Chru\'{s}ci\'{n}ski, J. Piilo, and W. T.
Strunz, Eternal non-Markovianity: from random unitary to Markov chain
realisations, Sci. Rep. \textbf{7}, 6379 (2017).

\bibitem{maximal} A. A. Budini, Maximally non-Markovian quantum dynamics
without environment-to-system backflow of information, Phys. Rev. A \textbf{%
97}, 052133 (2018).

\bibitem{pekola} B. Donvil, P. Muratore-Ginanneschi, and J. P. Pekola,
Hybrid master equation for calorimetric measurements, Phys. Rev. A \textbf{99%
}, 042127 (2019).

\bibitem{swf} H. P. Breuer, B. Kappler, and F. Petruccione, Stochastic
wave-function method for non-Markovian quantum master equations, Phys. Rev.
A \textbf{59}, 1633 (1999); H. P. Breuer, Genuine quantum trajectories for
non-Markovian processes, Phys. Rev. A \textbf{70}, 012106 (2004).

\bibitem{hush} M. R. Hush, I. Lesanovsky, and J. P. Garrahan, Generic map
from non-Lindblad to Lindblad master equations, Phys. Rev. A \textbf{91},
032113 (2015).

\bibitem{embedding} A. A. Budini, Embedding non-Markovian quantum
collisional models into bipartite Markovian dynamics, Phys. Rev. A \textbf{88%
}, 032115 (2013); A. A. Budini and P. Grigolini, Non-Markovian nonstationary
completely positive open-quantum-system dynamics, Phys. Rev. A \textbf{80},
022103 (2009).

\bibitem{collisionVacchini} B. Vacchini, Non-Markovian master equations from
piecewise dynamics, Phys. Rev. A \textbf{87}, 030101(R) (2013).

\bibitem{palmaMultipartito} V. Giovannetti and G. M. Palma, Master Equations
for Correlated Quantum Channels, Phys. Rev. Lett. \textbf{108}, 040401
(2012).

\bibitem{brasil} N. K. Bernardes, A. R. R. Carvalho, C. H. Monken, and M. F.
Santos, Environmental correlations and Markovian to non-Markovian
transitions in collisional models, Phys. Rev. A \textbf{90}, 032111 (2014).

\bibitem{ciccarello} F. Ciccarello, G. M. Palma, and V. Giovannetti,
Collision-model-based approach to non-Markovian quantum dynamics, Phys. Rev.
A \textbf{87}, 040103(R) (2013); S. Lorenzo, F. Ciccarello, and G. M. Palma,
Class of exact memory-kernel master equations, Phys. Rev. A \textbf{93},
052111 (2016); S. Lorenzo, F. Ciccarello, and G. M. Palma, Composite quantum
collision models, Phys. Rev. A \textbf{96}, 032107 (2017).

\bibitem{strunz} S. Kretschmer, K. Luoma, and W. T. Strunz, Collision model
for non-Markovian quantum dynamics, Phys. Rev. A \textbf{94}, 012106 (2016).

\bibitem{portugal} B. \c{C}akmak, M. Pezzutto, M. Paternostro, and \"{O}. E.
M\"{u}stecapl\i oglu, Non-Markovianity, coherence, and system-environment
correlations in a long-range collision model, Phys. Rev. A \textbf{96},
022109 (2017).

\bibitem{brasilCollisional} R. Ramirez Camasca and G. T. Landi, Memory
kernel and divisibility of Gaussian collisional models, Phys. Rev. A \textbf{%
103}, 022202 (2021).

\bibitem{Semi} H. P. Breuer and B. Vacchini, Quantum Semi-Markov Processes,
Phys. Rev. Lett. \textbf{101}, 140402 (2008); H. P. Breuer and B. Vacchini,
Structure of completely positive quantum master equations with memory
kernel, Phys. Rev. E \textbf{79}, 041147 (2009); B. Vacchini, Generalized
Master Equations Leading to Completely Positive Dynamics, Phys. Rev. Lett. 
\textbf{117}, 230401 (2016).

\bibitem{andrez} D. Chru\'{s}ci\'{n}ski and A. Kossakowski, Generalized
semi-Markov quantum evolution, Phys. Rev. A \textbf{95}, 042131 (2017); D.
Chru\'{s}ci\'{n}ski and A. Kossakowski, From Markovian semigroup to
non-Markovian quantum evolution, Euro Phys. Lett. \textbf{97}, 20005 (2012).

\bibitem{wudarski} D. Chru\'{s}ci\'{n}ski and F. A. Wudarski, Non-Markovian
random unitary qubit dynamics, Phys. Lett. A \textbf{377}, 1425 (2013); D.
Chru\'{s}ci\'{n}ski and F. A. Wudarski, Non-Markovianity degree for random
unitary evolution, Phys. Rev. A \textbf{91}, 012104 (2015).

\bibitem{Polonia} F. A. Wudarski, P. Nalezyty, G. Sarbicki, and D. Chru\'{s}%
ci\'{n}ski, Admissible memory kernels for random unitary qubit evolution,
Phys. Rev. A \textbf{91}, 042105 (2015); F. A. Wudarski and D. Chru\'{s}ci%
\'{n}ski, Markovian semigroup from non-Markovian evolutions, Phys. Rev. A 
\textbf{93}, 042120 (2016); D. Chru\'{s}ci\'{n}ski and K. Siudzi\'{n}ska,
Generalized Pauli channels and a class of non-Markovian quantum evolution,
Phys. Rev. A \textbf{94}, 022118 (2016); K. Siudzi\'{n}ska and D. Chru\'{s}ci%
\'{n}ski, Memory kernel approach to generalized Pauli channels: Markovian,
semi-Markov, and beyond, Phys. Rev. A \textbf{96}, 022129 (2017); K. Siudzi%
\'{n}ska, Markovian semigroup from mixing noninvertible dynamical maps,
Phys. Rev. A \textbf{103}, 022605 (2021).

\bibitem{TwoQubits} E. Ferraro, M. Scala, R. Migliore, and A. Napoli,
Non-Markovian dissipative dynamics of two coupled qubits in independent
reservoirs: Comparison between exact solutions and master-equation
approaches, Phys. Rev. A \textbf{80}, 042112 (2009).

\bibitem{exactDecayTLS} B. Vacchini and H. P. Breuer, Exact master equations
for the non-Markovian decay of a qubit, Phys. Rev. A \textbf{81}, 042103
(2010).

\bibitem{DivergingRatesJCModel} D. Maldonado-Mundo, P. \"{O}hberg, B. W.
Lovett, and E. Andersson, Investigating the generality of time-local master
equations, Phys. Rev. A \textbf{86}, 042107 (2012).

\bibitem{additivity} J. Lankinen, H. Lyyra, B. Sokolov, J. Teittinen, B.
Ziaei, and S. Maniscalco, Complete positivity, finite-temperature effects,
and additivity of noise for time-local qubit dynamics, Phys. Rev. A \textbf{%
93}, 052103 (2016).

\bibitem{ferialdi} L. Ferialdi, Exact non-Markovian master equation for the
spin-boson and Jaynes-Cummings models, Phys. Rev. A \textbf{95}, 020101(R)
(2017).

\bibitem{exactChina} H. Z. Shen, D. X. Li, Shi-Lei Su, Y. H. Zhou, and X. X.
Yi, Exact non-Markovian dynamics of qubits coupled to two interacting
environments, Phys. Rev. A \textbf{96}, 033805 (2017).

\bibitem{plenio} D. Tamascelli, A. Smirne, S. F. Huelga, and M. B. Plenio,
Nonperturbative Treatment of non-Markovian Dynamics of Open Quantum Systems,
Phys. Rev. Lett. \textbf{120}, 030402 (2018).

\bibitem{smirne} N. Megier, A. Smirne, and B. Vacchini, The interplay
between local and non-local master equations: exact and approximated
dynamics, New J. Phys. \textbf{22}, 083011 (2020).

\bibitem{deltaCorrelated} D. Burgarth, P. Facchi , M. Ligab\`{o}, and D.
Lonigro, Hidden non-Markovianity in open quantum systems, Phys. Rev. A 
\textbf{103}, 012203 (2021).

\bibitem{modi} F. A. Pollock, C. Rodr\'{\i}guez-Rosario, T. Frauenheim, M.
Paternostro, and K. Modi, Operational Markov Condition for Quantum
Processes, Phys. Rev. Lett. \textbf{120}, 040405 (2018).

\bibitem{budiniCPF} A. A. Budini, Quantum Non-Markovian Processes Break
Conditional Past-Future Independence, Phys. Rev. Lett. \textbf{121}, 240401
(2018); A. A. Budini, Conditional past-future correlation induced by
non-Markovian dephasing reservoirs, Phys. Rev. A \textbf{99}, 052125 (2019).

\bibitem{BIF} A. A. Budini, Detection of bidirectional system-environment
information exchanges, Phys. Rev. A \textbf{103}, 012221 (2021).

\bibitem{canonicalCresser} M. J. W. Hall, J. D. Cresser, L. Li, and E.
Andersson, Canonical form of master equations and characterization of
non-Markovianity, Phys. Rev. A \textbf{89}, 042120 (2014).

\bibitem{eigen} H. J. Briegel and B. G. Englert, Quantum optical master
equations: The use of damping bases, Phys. Ref. A \textbf{47}, 3311 (1993);
S. M. Barnetts and S. Stenholm, Spectral decomposition of the Lindblad
operator, J. Mod. Optics \textbf{47}, 2869 (2000).

\bibitem{BreuerFirst} H. P. Breuer, E. M. Laine, and J. Piilo, Measure for
the Degree of Non-Markovian Behavior of Quantum Processes in Open Systems,
Phys. Rev. Lett. \textbf{103}, 210401 (2009).

\bibitem{DarioSabrina} D. Chru\'{s}ci\'{n}ski and S. Maniscalco, Degree of
Non-Markovianity of Quantum Evolution, Phys. Rev. Lett. \textbf{112}, 120404
(2014).

%\bibitem{buchleitner} C. M. Kropf, C. Gneiting, and A. Buchleitner,
%Effective Dynamics of Disordered Quantum Systems, Phys. Rev. X \textbf{6},
%031023 (2016); C. Gneiting, F. R. Anger, and A. Buchleitner, Incoherent
%ensemble dynamics in disordered systems, Phys. Rev. A \textbf{93}, 032139 (2016).

%\bibitem{ciracR} B. Paredes, F. Verstraete, and J. I. Cirac, Exploiting
%Quantum Parallelism to Simulate Quantum Random Many-Body Systems, Phys. Rev.
%Lett. \textbf{95}, 140501 (2005).

\bibitem{tanh} We used the equality $(d/dt)\ln
[pe^{at}+qe^{bt}]=(a+b)/2+\Delta \tanh [t\Delta +\zeta ],$ where $\Delta
=(a-b)/2$ and $\zeta =(1/2)\ln (p/q).$
\end{thebibliography}
\end{document}